\documentclass{article}

\PassOptionsToPackage{numbers, compress}{natbib}

\usepackage[preprint]{neurips_2026}
\usepackage{wrapfig}
\usepackage{graphicx}
\usepackage{algorithm}
\usepackage{algorithmic}
\usepackage{float}
\usepackage{subcaption}
\usepackage{rotating}
\usepackage{caption}

\usepackage{color, colortbl}
\definecolor{CustomGray}{gray}{0.90}

\usepackage[utf8]{inputenc} 
\usepackage[T1]{fontenc}    
\usepackage{hyperref}       
\usepackage{url}            
\usepackage{booktabs}       
\usepackage{amsfonts}       
\usepackage{nicefrac}       
\usepackage{microtype}      
\usepackage{xcolor}         

\title{AnalogToBi: Device-Level Analog Circuit Topology Generation via Bipartite Graph and Grammar Guided Decoding}

\author{
Seungmin Kim$^{1}$ \quad
Mingun Kim$^{2}$ \quad
Yuna Lee$^{2}$ \quad
Yulhwa Kim$^{3}$\thanks{Corresponding author}
\\[0.5em]
$^{1}$Department of Semiconductor Display Engineering, Sungkyunkwan University, Republic of Korea
\\
$^{2}$Department of Electrical and Computer Engineering, Sungkyunkwan University, Republic of Korea
\\
$^{3}$Department of Semiconductor Systems Engineering, Sungkyunkwan University, Republic of Korea
\\[0.5em]
\texttt{\{seungmin.kim, kmgun110, yuna6548, yulhwakim\}@skku.edu}
}

\begin{document}

\maketitle

\begin{abstract}
Analog circuit design remains highly dependent on expert knowledge due to the complexity of device-level interactions and topology design. Recent transformer-based approaches for device-level topology generation have shown promise, yet they suffer from low electrical validity without human-in-the-loop (HITL) training and severe memorization caused by sequence-based circuit representations.
In this work, we propose \textbf{AnalogToBi}, a framework for device-level analog circuit topology generation. AnalogToBi introduces \emph{circuit-type conditioning} for categorizing heterogeneous multi-type topology datasets, \emph{device renaming augmentation} to mitigate memorization, \emph{a bipartite graph representation} for improved structural generalization, and \emph{grammar-guided decoding} to enforce structural validity during bipartite graph generation.
Experimental results demonstrate that AnalogToBi achieves high validity and novelty without HITL training while effectively avoiding memorization of training topologies. Our code is available at \url{https://github.com/Seungmin0825/AnalogToBi}.
\end{abstract}

\section{Introduction}
Analog circuits form the backbone of modern electronic systems, yet their design remains highly dependent on expert knowledge~\cite{liakos2025analogml}.
Analog circuit design typically involves determining a circuit topology that defines the core functionality of the circuit, followed by device sizing to achieve desired operating conditions~\cite{liu2025eesizer}.
This process requires deep understanding of device-level behavior and complex inter-device interactions, resulting in strong reliance on experienced analog designers.
However, the limited availability of such expertise across the semiconductor industry has motivated growing interest in analog design automation.

Early analog design automation research primarily focused on device sizing optimization~\cite{settaluri2020autockt, cao2022domain, wang2020gcnrl, lyu2017bo}.
While effective for parameter refinement, these approaches assume that the circuit topology is given.
Since circuit topology fundamentally determines analog circuit functionality, this assumption limits their applicability.
Therefore, topology generation is a critical step toward fully automated analog circuit design.

Automating circuit topology design is substantially more challenging than device sizing optimization, as it requires modeling complex structural and electrical relationships among devices.
As a result, early analog design automation research largely avoided explicit topology generation despite its importance.
Recent advances in transformer-based models, including large language models (LLMs), have renewed interest in this problem by enabling improved modeling of contextual information~\cite{lai2025analogcoder, chang2024lamagic, gao2025analoggenie}.
However, existing approaches still suffer from important limitations, including low electrical validity, memorization of training topologies, and reliance on HITL optimization.
To address these challenges, we propose AnalogToBi, a novel framework for device-level analog circuit topology generation.

\begin{figure}
    \centering
    \includegraphics[width=\linewidth]{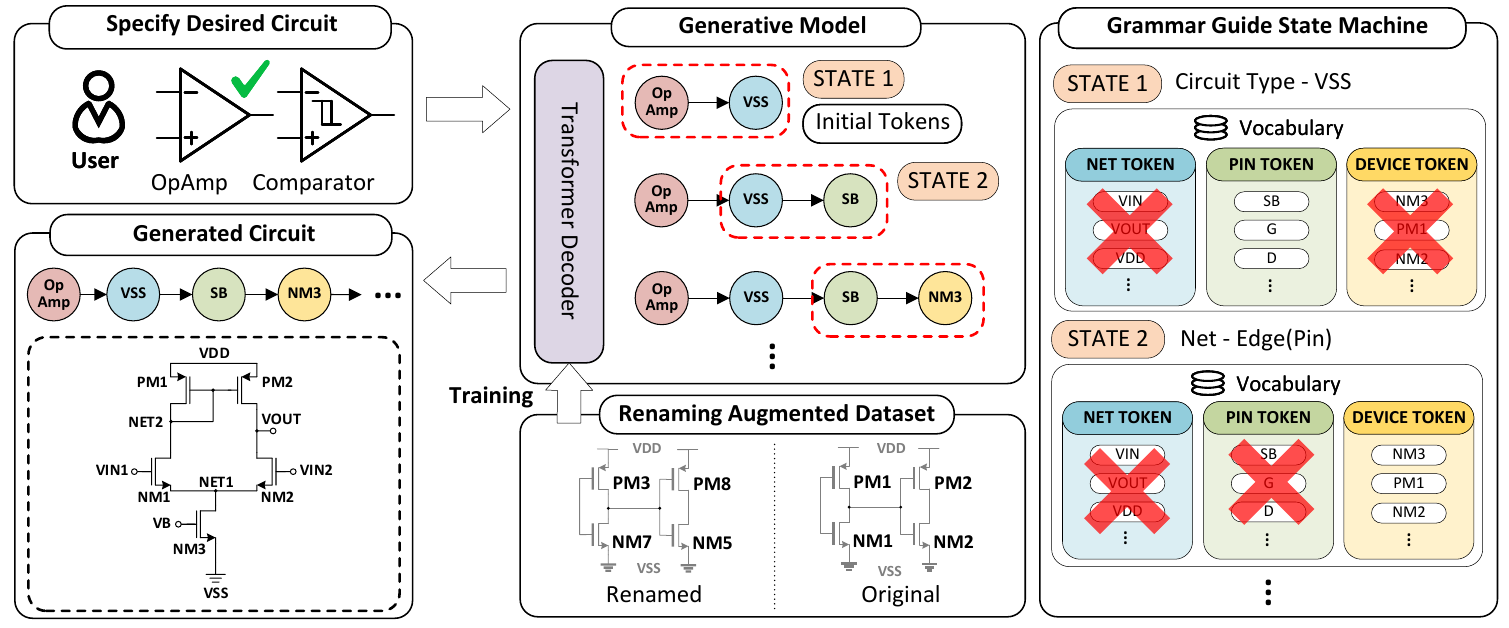}
    \caption{
    Overview of the proposed AnalogToBi framework. Given a circuit type, a Transformer decoder trained with renaming augmentation generates a device-level circuit sequence under grammar-guided decoding, which is converted into a bipartite device–net graph and netlist.
    }
\end{figure}

\section{Related Work}
\label{sec:related}

This section reviews prior work on the automation of analog circuit topology design.
Previous approaches can be categorized into four classes: topology search, topology refinement, behavior-level topology generation, and device-level topology generation.
The following subsections examine the key characteristics of each category and conclude by identifying the limitations of prior work that motivate the proposed approach.


\subsection{Topology Search and Topology Refinement}

\textbf{Topology Search.}
Previous works in this category aim to identify topologies that satisfy user-specified operational goals from predefined candidate sets manually designed based on prior analog circuit knowledge~\cite{veselinovic1995flexible, gerlach2015toposelect, poddar2024datetopo, mehradfar2025falcon}.
For example, Falcon~\cite{mehradfar2025falcon} selects the most suitable topology for a given target specification.
However, because these methods operate only within predefined candidate sets, they cannot generate topologies beyond those explicitly included.

\textbf{Topology Refinement.}
Recent studies have explored the use of pre-trained LLMs for topology refinement, motivated by their strong capabilities in reasoning, code generation, and tool-integrated agentic workflows~\cite{liu2024ladac, bhandari2024masala}.
Representative works include AnalogCoder~\cite{lai2025analogcoder, lai2025analogcoderpro}, Artisan~\cite{chen2024artisan}, AnalogExpert~\cite{zhang2025analogxpert}, and AnalogSeeker~\cite{chen2025analogseeker}.
In these approaches, users provide an initial circuit topology, typically represented in a code-like format such as a SPICE netlist, allowing pre-trained LLMs to interpret circuit structures.
To further improve domain understanding, AnalogExpert~\cite{zhang2025analogxpert} additionally fine-tunes pre-trained LLMs for analog design tasks.
The LLMs then act as agents that iteratively refine the topology using circuit simulation feedback.
While effective for design refinement, these approaches remain highly dependent on the quality of the initial topology, limiting their ability to systematically explore novel circuit structures.

\subsection{Behavioral-Level Topology Generation}
Early topology generation approaches primarily adopted behavior-level topology generation strategies~\cite{zhao2019grammar, lu2022graphembed, dong2023cktgnn}.
In these approaches, key analog circuit components are abstracted as behavior-level building blocks whose types and internal designs are predefined.
Topology generation is then performed by composing and connecting these abstract blocks.
For example, CktGNN~\cite{dong2023cktgnn} treats single-stage operational amplifiers (OpAmps), along with resistors and capacitors, as the basic building blocks for topology generation.
While this formulation enables the generation of diverse circuit structures by composing these basic units, it inherently limits design flexibility, as the internal device-level realizations of single-stage OpAmps are fixed and circuits that cannot be expressed as compositions of abstract blocks cannot be generated.
As a result, behavior-level topology generation methods are unable to sufficiently capture the rich design space of device-level circuit topologies.

\subsection{Device-Level Topology Generation}
\label{sec:related_device_level}
More recently, active research has explored device-level topology generation to enable fully flexible analog circuit design~\cite{chang2024lamagic, chang2025lamagic2, gao2025analoggenie, gao2025analoggenielite, li2025analogfed}.
These approaches leverage transformer-based models to directly generate device-level circuit representations. 
However, because the sequential representations of analog circuits differ substantially from natural language, such methods typically train dedicated models rather than relying on pre-trained language models.
LaMAGIC~\cite{chang2024lamagic, chang2025lamagic2}, a pioneering work in this category, proposes training a new model for device-level topology generation.
While effective within its target domain, it is limited to a single circuit type, specifically power converters.

AnalogGenie~\cite{gao2025analoggenie, gao2025analoggenielite} extends this line of work by training a single model on a dataset covering multiple circuit types, enabling broader topology generation.
However, to train a high-quality topology generation model that produces a high ratio of electrically valid circuits, AnalogGenie relies on HITL reinforcement learning, where expert designers evaluate generated circuits and provide feedback during training.
In this process, the criteria for quality evaluation are not explicitly defined and are largely dependent on individual expert judgment, thereby limiting reproducibility and scalability.
Moreover, its device-pin-level graph representation encourages memorization rather than structural generalization.


In summary, enabling fully flexible analog circuit topology design ultimately requires device-level topology generation.
To overcome the limitations of existing approaches, there is a clear need for a topology generation framework that ensures high electrical validity without HITL training and promotes topological novelty without memorization.


\section{Proposed Method}

We propose AnalogToBi, a novel device-level analog circuit topology generation framework.
Similar to recent approaches~\cite{gao2025analoggenie, gao2025analoggenielite}, AnalogToBi employs a transformer-based model dedicated to analog circuit topology generation. 
To overcome the limitations of prior work, AnalogToBi introduces four key design elements, which are described in detail in the following subsections.

\subsection{Circuit-Type Conditioning and Data Preprocessing}
\label{sec:circuit_type}

\begin{wrapfigure}{r}{0.52\textwidth}
    \centering
    \vspace{-4pt}
    \includegraphics[width=0.50\textwidth]{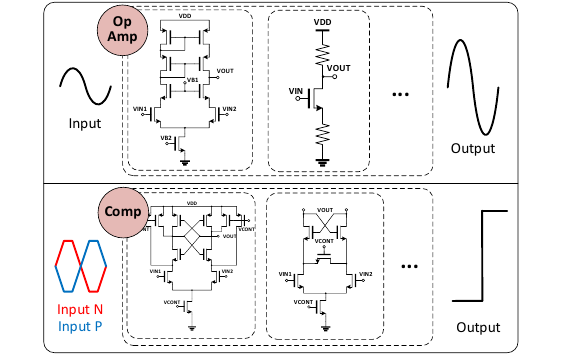}
    \caption{
    Example analog circuit topologies for two circuit types: OpAmp and comparator (Comp).
    }
    \label{fig:circuit_type}
    \vspace{-6pt}
\end{wrapfigure}

As discussed in Section~\ref{sec:related_device_level}, AnalogGenie~\cite{gao2025analoggenie, gao2025analoggenielite} is trained on a dataset covering multiple circuit types to support broader topology generation.
However, circuits from different functional categories exhibit fundamentally different topological structures. From the perspective of generative modeling, such an unstructured dataset induces a multimodal distribution over circuit topologies.
For example, as shown in Fig.~\ref{fig:circuit_type}, OpAmps typically employ a differential pair with a current mirror load, optionally augmented with cascode stages or extended to two-stage configurations.
In contrast, comparators often rely on latch-based structures or pre-amplifier plus high-gain stage architectures.
These examples highlight that circuit topology is tightly coupled with functionality, and that different circuit types follow distinct structural rules.



Training a generative model on such heterogeneous data without explicit conditioning can lead to undesirable behavior.
In particular, the model may interpolate between distinct modes, resulting in physically invalid circuits that violate underlying design principles.
To address this, we use a preprocessed training dataset in which circuits are categorized into functional types, thereby providing conditional constraints that guide the model toward electrically valid topologies.

Following widely adopted functional categorizations in analog design, such as OpAmps and comparators, the 2{,}165 raw netlists from the AnalogGenie dataset are categorized into 15 circuit types (see Appendix~\ref{app:dataset} for details).
We further incorporate a circuit-type token into the model vocabulary, which is provided as the first input token to explicitly condition the generation process.

However, naively introducing data preprocessing and circuit-type tokens into the AnalogGenie framework exacerbates the memorization problem.
This primarily stems from data sparsity, as partitioning the dataset into multiple circuit types reduces the number of training samples available for each category.
With fewer examples per type, the model tends to overfit to typical patterns rather than generalizing to diverse topologies.
To mitigate this issue, we introduce a renaming data augmentation strategy in the following section.

\subsection{Data Augmentation with Device Renaming}
\label{section:augmentation}

Transformer-based circuit generation models operate on sequential representations, 
requiring each circuit to be converted into a token sequence.
Each circuit is first transformed into a graph representation and then serialized 
into a sequence based on a graph traversal order.
Because different traversal orders produce different sequences for the same circuit 
graph, previous work~\cite{gao2025analoggenie, gao2025analoggenielite} introduces 
a data augmentation strategy that generates multiple sequence variants by applying 
different traversal orderings.
While this traversal-order augmentation increases the number of training sequences, 
it does not prevent the model from memorizing sequence patterns, as the original 
device indices from the raw dataset are repeatedly exposed across augmented variants.
To address this limitation, we introduce an additional \textit{device renaming 
augmentation} that randomly reassigns device identifiers across training samples.
This augmentation exploits the fact that device identifiers encode instance names 
rather than functional behavior: renaming a device token such as \texttt{NMOS1} 
to \texttt{NMOS7} does not alter the electrical functionality or connectivity of 
the circuit (see appendix ~\ref{app:renaming} for details).
By randomizing device indices, this augmentation disrupts index-based pattern 
memorization while preserving electrical equivalence, thereby encouraging the model 
to learn structurally grounded representations rather than surface-level token patterns.


Our evaluation shows that device renaming augmentation effectively alleviates memorization, even under circuit-type conditioning.
However, when applied to the original AnalogGenie framework, the improvement in novelty comes at the cost of degraded validity (a detailed quantitative analysis is provided in Section~\ref{sec:ablation}).
This trade-off stems from the underlying circuit representation used in AnalogGenie.
AnalogGenie adopts a device–pin level graph representation for analog circuit modeling (Fig.~\ref{fig:vocab}-left), in which each pin of a device is assigned a distinct token.
It supports circuit generation with up to 35 NMOS and 35 PMOS devices (e.g., NM1, $\ldots$, NM35 and PM1, $\ldots$, PM35), which are further expanded into pin-level tokens such as NM1S, NM1G, NM1D, and NM1B, where S, G, D, and B denote source, gate, drain, and body, respectively.
This results in a large and fixed token space with $70 \times 4$ distinct pin-specific tokens.
While this representation explicitly encodes pin-level connectivity in a sequential format, it effectively reduces circuit generation to an ordering problem over a fixed token set.
Instead of learning the underlying connectivity structure, the model is encouraged to learn how to arrange a predefined set of tokens into valid sequences.
As a result, the model tends to memorize token-level patterns rather than learning generalizable structural relationships.
Consequently, validity in the original framework is largely maintained through memorization, and disrupting this memorization via augmentation leads to a degradation in validity.
This reveals a fundamental limitation of the representation, where validity and novelty are inherently entangled through memorization.

\subsection{Bipartite Graph Circuit Representation}
\label{sec:proposed_bipartite}

\begin{figure}[t]
  \centering
  \includegraphics[width=\linewidth]{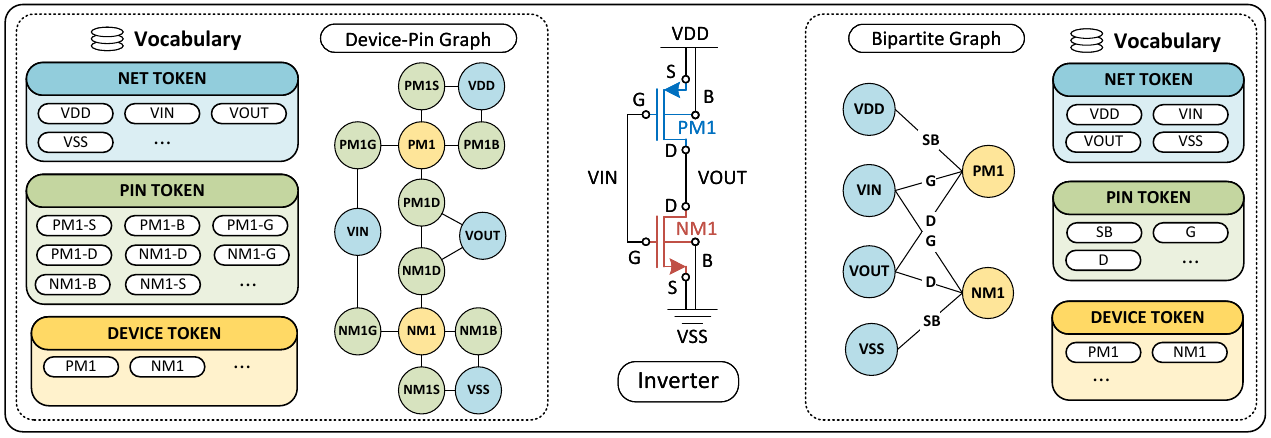}
    \caption{Comparison between conventional device–pin level representation~\cite{gao2025analoggenie} and the proposed bipartite graph representation. The device–pin representation (left) treats each device pin as a separate token, while the bipartite representation (right) models devices and nets as distinct node types with pin semantics encoded as typed edges.}
  \label{fig:vocab}
  \vspace{-8pt}
\end{figure}

To encourage the model to learn generalizable topology patterns rather than memorizing training sequences, we propose a more flexible representation that decouples device-specific identifiers from functional connectivity.
To this end, we leverage the inherent structural components of analog circuits, namely devices, nets, and pins (Fig.~\ref{fig:vocab}-right).
In analog circuit design, devices perform electrical functions, nets represent electrical connection points, and pins define the interfaces through which devices connect to nets.
A key observation is that a device can connect to a net only through its pins, and conversely, a net is always connected to one or more devices.
This mutual dependency implies that any analog circuit can be naturally expressed as a graph consisting of two disjoint entity types, devices and nets, with connectivity mediated exclusively by pins.
Based on this observation, we represent each analog circuit as a bipartite graph composed of a device node set and a net node set.
Edges between device nodes and net nodes correspond to pin-level connections and indicate which pin of a device is connected to which net.
Importantly, pins are not treated as independent entities. Instead, they are modeled as intrinsic attributes of devices that define the type of connection to a net.
This formulation eliminates the need to define device-specific pin tokens as standalone nodes.
To process this bipartite representation with the proposed model, we categorize tokens into device-type tokens (e.g., NM1, PM1), net-type tokens (e.g., VIN1, NET1), and pin-type tokens (e.g., S, D).
The bipartite graph is then serialized into a sequence using a graph traversal strategy detailed in appendix ~\ref{app:augmentation}.
In this representation, the position and function of each pin are resolved relationally through adjacent device and net tokens.
This contextual interpretation of pin semantics encourages the model to reason over structural context, since identifying the role of a pin requires understanding its neighboring tokens and their relationships.


However, the proposed bipartite circuit representation imposes a specific ordering constraint during token generation, in which pin-type tokens must appear between device-type and net-type tokens.
If the model fails to internalize this constraint, electrically invalid configurations such as floating nodes or short circuits may arise during generation.
Given the limited availability of device-level analog circuit data, it is challenging for the model to reliably learn and enforce this ordering rule through training alone.
This limitation motivates the grammar-guided decoding strategy introduced in the next section.

\subsection{Grammar-Guided Decoding for Bipartite Graph Representation}
\label{sec:grammar_guided_decoding}

\begin{wrapfigure}{l}{0.52\textwidth}
    \centering
    \vspace{-8pt}
    \includegraphics[width=0.50\textwidth]{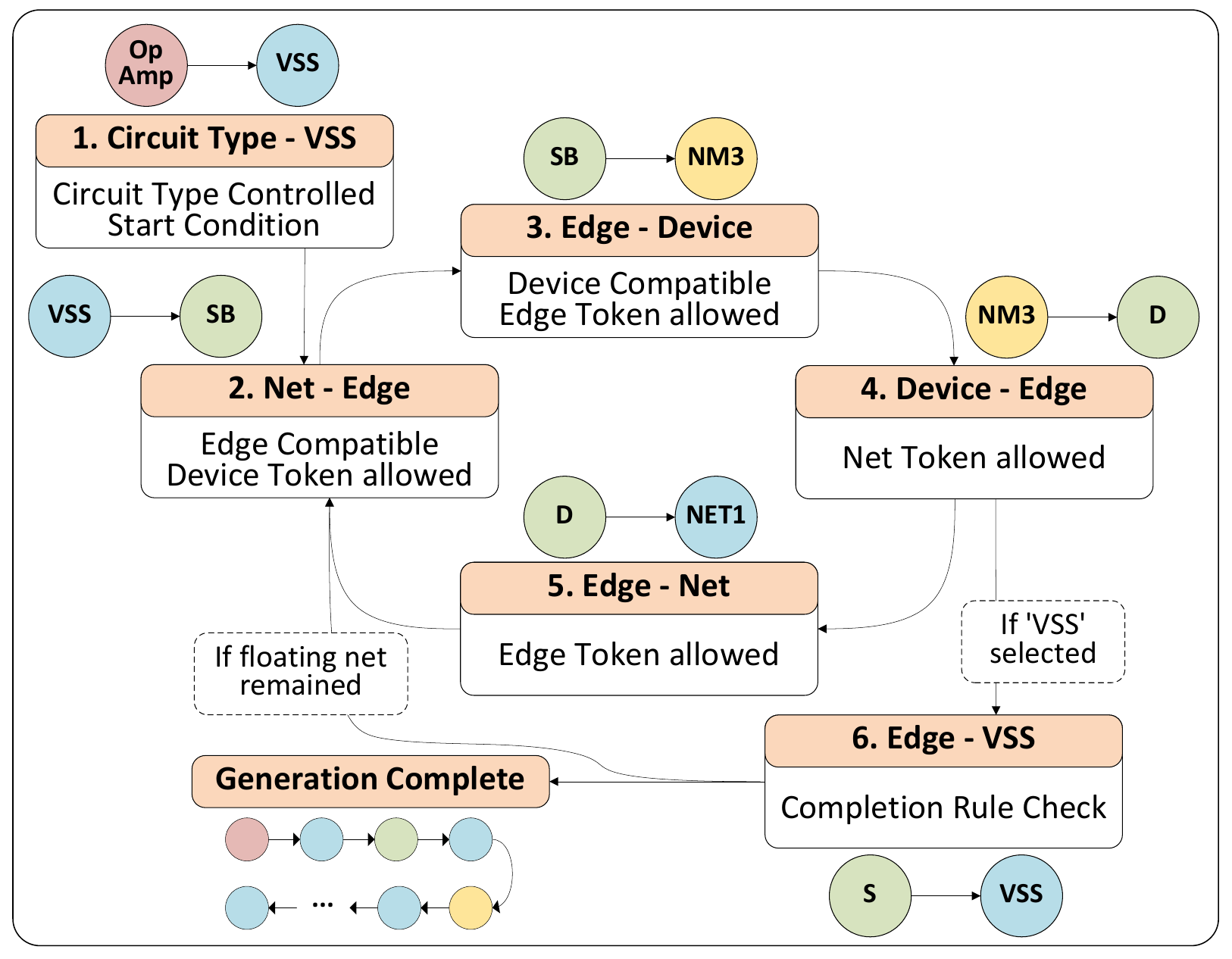}
    \caption{
    Grammar-guided decoding state machine.
    }
    \label{fig:grammar}
    \vspace{-0pt}
\end{wrapfigure}

To enable effective topology generation based on the bipartite graph representation under limited data availability, we introduce grammar-guided decoding.
Following the paradigm of grammar-constrained decoding in LLMs~\cite{geng2023grammar, raspanti2025grammar}, our approach enforces circuit-level electrical and structural constraints directly during the generation of token sequences, as illustrated in Fig.~\ref{fig:grammar}.

We formalize the structural regularities induced by the bipartite graph circuit representation as an explicit grammar.
Circuit generation is modeled as a state-transition process, where each state is defined by the combination of the token generated at the current decoding step $T$ and the token generated at the previous step $T-1$.

Rather than operating on individual token identities, states are defined in terms of token categories, namely circuit-type tokens, device-type tokens, net-type tokens, and pin-type tokens.
Circuit-type tokens are provided by the user to specify the target circuit functionality and are used solely to initialize the generation process.
These tokens are not generated during decoding and are therefore masked out for all subsequent generation steps.

Given the current state, the grammar restricts the set of admissible next tokens by masking invalid transitions, enforcing a structured device–pin-net transition pattern.
For example, if a device-type token is generated at step $T$, then at step $T+1$ all tokens except pin-type tokens are masked, forcing the model to generate a pin-type token.
By excluding invalid transitions from the generation trajectory, this state-machine-based decoding prevents electrically invalid configurations from being generated.
Finally, sequence termination is permitted only when all device pins are fully connected and no internal net remains floating, thereby guaranteeing the electrical completeness of the generated circuit.
More details on the specific grammar adopted in AnalogToBi are provided in Appendix~\ref{app:grammar}.


\section{Experiments}
\label{section:experiment}

\subsection{Experimental Setup}
\label{section:experimental}
\textbf{Dataset.}
We evaluate the proposed AnalogToBi using the publicly released dataset of 2,165 analog circuits represented as SPICE-style netlists~\cite{gao2025analoggenie}.
To ensure stable model training, we use the functionally categorized dataset described in Section~\ref{sec:circuit_type}.
For each circuit category, 90\% of the circuits are used for training, while the remaining 10\% are reserved for validation.
Each circuit is first converted into a graph representation and subsequently serialized into a sequence.

We apply two forms of data augmentation:
(1) the sequence augmentation strategy proposed in AnalogGenie, which generates multiple sequence variants for each circuit using different graph traversal orderings, and
(2) the proposed device renaming augmentation (Section~\ref{section:augmentation}), which renames device tokens without altering circuit functionality, thereby increasing data diversity while preserving electrical equivalence.
As a result of these augmentations, the effective training dataset is expanded from 2,165 to 397,515 sequences.

\textbf{Baselines.}
We compare AnalogToBi with recent approaches for device-level topology generation, including LaMAGIC~\cite{chang2024lamagic}, and AnalogGenie(-Lite)~\cite{gao2025analoggenie, gao2025analoggenielite}.
These methods differ from AnalogToBi in terms of circuit representation, generation capability, and scalability, as discussed in Section~\ref{sec:related}.
LaMAGIC focuses on generating power converter circuits within a single circuit domain, resulting in highly limited topology exploration.
AnalogGenie serves as the strongest baseline for our task, as it supports the generation of multiple circuit types. However, it requires HITL feedback to achieve high electrical validity and suffers from severe memorization of training circuits.
For all baselines, we follow the original implementations and experimental settings to reproduce the reported results.


\textbf{Implementation Details.}
To ensure a fair comparison with the strong baseline AnalogGenie, we align the model architecture used for AnalogToBi, thereby isolating the impact of the proposed circuit representation and grammar-guided decoding mechanisms.
The decoder-only transformer has an embedding dimension of 384, 6 attention heads, and 6 transformer layers, resulting in approximately 11.3M parameters.
Note that the model size of AnalogToBi is slightly smaller than that of AnalogGenie (11.8M parameters), due to the more compact token vocabulary and corresponding embedding layers enabled by the proposed bipartite representation, as discussed in Section~\ref{sec:proposed_bipartite}.
The maximum context length is fixed to 1024 tokens, and a dropout rate of 0.2 is applied throughout the network.
Training is performed using AdamW with a learning rate of $3\times10^{-4}$ and a batch size of 64 for 100{,}000 iterations.
The entire training process takes 22.8 hours on an NVIDIA RTX 5880 GPU.
During inference, circuits are generated autoregressively using multinomial sampling with a temperature of $\tau=0.7$.

\textbf{Metrics.}
To properly evaluate the ability of the circuit generation model to generate structurally correct and novel circuit topologies, we assess performance using five metrics: Validity, Novelty, Valid \& Novel, Exact sequence matching (Exact.), Graph Edit Distance (GED).
\textbf{Validity} measures the fraction of generated circuits that are electrically valid, such that they can be converted into SPICE netlists without errors.
While this metric alone does not guarantee that the generated circuits function properly, previous works on device-level topology generation has primarily adopted this level of validity evaluation to assess topology quality~\cite{gao2025analoggenie, gao2025analoggenielite}.
For fair comparison, we follow the same evaluation protocol.
We note that improving the ratio of fully correctly operating generated circuits remains an important future direction for device-level topology generation research.
\textbf{Novelty} measures the fraction of generated circuits whose topologies are not observed in the training set.
We evaluate novelty using graph isomorphism, where a generated circuit is considered novel if its topology is not isomorphic to any training circuit.
Although prior works~\cite{gao2025analoggenie, gao2025analoggenielite} also reported novelty metrics, their exact evaluation methodology was not explicitly described.
Therefore, the novelty values reported in our work may not be directly comparable to those reported in previous studies.
\textbf{Valid \& Novel} measures the proportion of generated circuits that are both electrically valid and novel. 
\textbf{Exact.} indicates the rate at which the transformer decoder model outputs the same sequence as the dataset used when learning.
A high Exact. score indicates that the model reproduces training samples verbatim rather than generating unseen structures.
Therefore, both Novelty and Exact. serve as indicators of the memorization tendency of the model.
\textbf{GED} denotes the average minimum graph edit distance between each generated topology and the topologies in the training dataset~\cite{bunke1998graph, papadimitriou2010web, shervashidze2011weisfeiler, fischer2017improved}.
It measures the structural difference between generated circuits and training samples.
A higher GED indicates that the model generates structurally more diverse circuits rather than producing minor variations of existing training topologies.


\begin{table*}
\centering
\caption{
Comparison of AnalogToBi with baselines.
\textit{Model(\#param)} denotes the model architecture and number of parameters.
\textit{HITL} indicates HITL involvement during training, and \textit{\#Type} denotes the number of circuit types in the training dataset.
All baseline results are taken from the original papers except for AnalogGenie*, which is reproduced using the officially released implementation.
Novelty marked with $^{+}$ corresponds to the value reported in the original AnalogGenie paper, whose evaluation methodology was not explicitly described.}
\label{tab:main_result_1_compare_with_prev}
\resizebox{\linewidth}{!}{
\begin{tabular}{l ccc c ccccc}
\toprule
 & \multicolumn{3}{c}{\bf Properties} && \multicolumn{5}{c}{\bf Evaluation Results}\\
 \cmidrule{2-4} \cmidrule{6-10}
 \bf \;Method	 & Model (\#param.) & HITL & \#Type  && Validity$\uparrow$ & Novelty$\uparrow$	& Val.\&Nov.$\uparrow$ & Exact.$\downarrow$ & GED$\uparrow$\\
\midrule \midrule 
\;LaMAGIC~\cite{chang2025lamagic2}					 & Flan-T5 (250M) & $\times$ & 1 &&96.0\%  &-- &-- &-- & \\
\;AnalogGenie~\cite{gao2025analoggenie}				 & Custom (11.8M) &$\times$ & 11 &&73.5\%  &98.9\%$^{+}$ &-- &-- & \\
\midrule
\;AnalogGenie~\cite{gao2025analoggenie}				 & Custom (11.8M) &\tiny $\bigcirc$ &11 &&93.2\%  &99.0\%$^{+}$ &-- &-- & \\
\;AnalogGenie-Lite~\cite{gao2025analoggenielite}\;\;		 & Custom (11.8M) &\tiny $\bigcirc$ &11 &&97.0\%  &99.8\%$^{+}$ &-- &-- & \\
\midrule
\;AnalogGenie*~\cite{gao2025analoggenie}		 		 & Custom (11.8M) &$\times$ &11 &&75.6\%  &50.4\% &26.0\% &39.6\% &209.5 \\
\rowcolor{CustomGray}
\;\bf AnalogToBi (Proposed)						 		 & \bf {Custom (11.3M)} & \bf $\times$ &\bf 15 &&\bf 97.8\%  &\bf 92.1\% &\bf 89.9\% &\bf 0\% &\bf 226.8 \\
\bottomrule
\end{tabular}
}
\end{table*}

\subsection{Main Results}
Table~\ref{tab:main_result_1_compare_with_prev} presents a comprehensive comparison between AnalogToBi and prior approaches in terms of both methodological properties and generation quality.
LaMAGIC achieves high validity (96.0\%) without HITL training or inference. However, it is limited to single-type topology generation, specifically power converters, which restricts its applicability to general analog circuit design.
AnalogGenie(-Lite) extends topology generation to multiple circuit types. However, without HITL-based expert feedback during training, its validity drops to 73.5\%. Achieving validity above 90\% requires HITL training, where expert designers evaluate generated circuits and provide feedback throughout optimization.
For fair comparison, we compare AnalogToBi against AnalogGenie without HITL, representing a fully automated training setting.
Because the novelty evaluation methodology of AnalogGenie was not explicitly described, we reproduced the original model using the officially released implementation and evaluated it using our metrics.
The reproduced model achieves a validity of 75.6\%, which is consistent with the originally reported result (73.5\%).
We denote this reproduced model as AnalogGenie* in Table~\ref{tab:main_result_1_compare_with_prev}.
AnalogToBi consistently outperforms AnalogGenie* across all evaluation metrics.
Specifically, AnalogToBi achieves a validity of 97.8\%, a novelty of 92.1\%, and an exact-match rate of 0\%, indicating that the model does not memorize training sequences.
Furthermore, the high Valid \& Novel ratio of 89.9\% demonstrates that the generated circuits remain highly novel even when considering only electrically valid samples.
This is important because novelty alone can be artificially increased by generating random invalid sequences.
The results therefore indicate that AnalogToBi successfully learns generalizable topology patterns and generates novel yet electrically valid circuits.
Moreover, AnalogToBi achieves a GED of 226.8, compared to 209.5 for AnalogGenie*, indicating stronger structural diversity and generation capability.
Overall, AnalogToBi generates electrically valid and structurally novel circuits without memorizing training data, all while eliminating the need for human intervention during training. 

\begin{table}[t]
\centering

\caption{
Ablation study of AnalogToBi evaluating the impact of four key design components.
}

\label{tab:result_ablation}

\resizebox{\linewidth}{!}{
\begin{tabular}{l ccc c cccc}
\toprule
 \multicolumn{4}{c}{\bf Options} && \multicolumn{4}{c}{\bf Evaluation Results}\\
\cmidrule{1-4} \cmidrule{6-9}
\;Representation \;\; & Type Token & Rename & Grammar&& Validity~$\uparrow$ & Novelty~$\uparrow$ & Val.\&Nov.~$\uparrow$ & Exact.~$\downarrow$ \\
\midrule \midrule

\;Device-pin & $\times$ & $\times$ & --
&& 75.6\% & 50.4\% & 26.0\% & 39.6\%     \\


\;Device-pin & \tiny $\bigcirc$ & $\times$  & --
&& 86.1\% & 29.2\% & 15.4\% & 57.5\%   \\

\;Device-pin & \tiny $\bigcirc$ & \tiny $\bigcirc$  & --
&& 55.6\% & 84.1\% & 39.7\% & 0\%   \\

\midrule

\;Bipartite& $\times$ & $\times$ & $\times$
&& 53.6\% & 61.0\% & 15.3\% & 0\%  \\





\;Bipartite& $\times$ & $\times$ & \tiny $\bigcirc$
&& 61.8\% & 57.3\% & 19.9\% & 0\%   \\


\;Bipartite  & \tiny $\bigcirc$ & $\times$ & \tiny $\bigcirc$
&& 97.8\% & 74.6\% & 72.4\% & 1.4\%   \\

\bf \;Bipartite& \tiny $\bigcirc$ & \tiny $\bigcirc$ & \tiny $\bigcirc$
&& \textbf{97.8\%} & \textbf{92.1\%} & \textbf{89.9\%} & \textbf{0\%}  \\

\bottomrule
\end{tabular}
}
\vspace{-3pt}
\end{table}

\subsection{Ablation Study}
\label{sec:ablation}

Table~\ref{tab:result_ablation} presents an ablation study of the proposed AnalogToBi framework, evaluating the impact of four key design components: circuit-type tokens, device renaming augmentation, circuit representation, and grammar-guided decoding.
Please note that the device–pin representation without circuit-type tokens or renaming augmentation corresponds to the original AnalogGenie framework.
When circuit-type tokens are introduced into the device–pin representation, validity improves from 75.6\% to 86.1\%. However, novelty drops from 50.4\% to 29.2\%, and Exact. increases from 39.6\% to 57.5\%.
This indicates that the device–pin representation tends to achieve higher validity by memorizing training samples rather than learning generalizable topology patterns.
Applying device renaming augmentation alleviates this memorization behavior, increasing novelty to 84.1\% while reducing Exact. to 0\%.
However, this improvement comes at the cost of validity, which drops significantly to 55.6\%.
These results suggest that the device–pin representation fundamentally lacks the ability to learn topology patterns without relying on memorization.

For the proposed bipartite graph representation, validity remains relatively low without grammar-guided decoding because learning the structural rules required for bipartite graph is challenging under limited training data.
By introducing grammar-guided decoding, validity improves by 8.2\%p, demonstrating that explicit structural constraints effectively assist the generation process.
Interestingly, without circuit-type tokens, the bipartite representation achieves lower validity than the device–pin representation while exhibiting substantially lower Exact. scores.
This indicates that the bipartite representation is less prone to memorization and instead attempts to learn topology-level structural patterns.
However, because the training dataset contains structurally heterogeneous circuit topologies without functional categorization, the bipartite representation struggles to achieve high validity.
Finally, when the bipartite representation is combined with circuit-type conditioning, validity increases substantially to 97.8\%.
Moreover, applying renaming augmentation on top of this configuration preserves validity while further increasing novelty to 92.1\%.
Overall, the results demonstrate that high validity, high novelty, and low memorization can be achieved when proposed design components are jointly integrated.

\begin{table*}[t]
\centering

\caption{
Breakdown of AnalogToBi generation results by circuit type.
(Mirror: current mirror; Comp: comparator; Mix: mixer; LDO: low-dropout regulator;
Oscil: oscillator; Filt: filter; BGR: bandgap reference; PAmp: power amplifier; VoltR: voltage regulator;
PConv: power converter; PLL: phase-locked loop; S Cap: switched-capacitor; A/D\_D/A: data converter)
}
\vspace{-3pt}
\label{tab:main_results_2_AnalogToBi_breakdown}

\resizebox{\linewidth}{!}{
\begin{tabular}{lccccccccccccccc|c}
\toprule
 & OpAmp & Mirror & Comp & Mix & LDO & Oscil & Filt & BGR & PAmp & VoltR & PConv & PLL & S Cap & A/D\_D/A & General & Avg. \\
\midrule
Validity (\%)       
& 97.6 & 99.4 & 97.1 & 99.3 & 95.6 & 97.7 & 98.9 & 97.7 & 97.7 & 98.5 & 96.2 & 97.8 & 95.9 & 99.6 & 98.1 & \textbf{97.8} \\

Novelty (\%)        
& 89.1 & 77.4 & 92.0 & 84.1 & 97.6 & 92.1 & 92.9 & 94.0 & 99.7 & 91.0 & 99.2 & 90.9 & 91.4 & 96.5 & 92.9 & \textbf{92.1} \\

Valid \& Novel (\%) 
& 86.7 & 76.8 & 89.1 & 83.5 & 93.2 & 89.8 & 91.8 & 91.7 & 97.4 & 89.5 & 95.4 & 88.7 & 87.3 & 96.1 & 91.0 & \textbf{89.9} \\

Type acc. (\%)  
& 95.8 & 92.4 & 86.4 & 99.6 & 100.0 & 96.7 & 93.6 & 100.0 & 95.3 & 98.5 & 100.0 & 89.6 & 95.2 & 95.1 & -- & \textbf{95.6} \\
\bottomrule
\end{tabular}
}
\end{table*}

\subsection{Additional Analysis on Circuit-Type Controllability}

As the training dataset is categorized by circuit type and circuit-type tokens are introduced during training, AnalogToBi has the potential to control the type of generated topologies through circuit-type conditioning.
To evaluate whether generated circuits match the intended category, we train a Graph Attention Network (GAT) classifier that predicts the circuit type of generated topologies.
We define type accuracy as the fraction of generated circuits whose predicted type matches the input circuit-type token.
Details of the GAT architecture and training procedure are provided in Appendix~\ref{app:gat}.
Table~\ref{tab:main_results_2_AnalogToBi_breakdown} reports the circuit-type-wise breakdown of generated circuits, where 1,000 samples are generated for each category.
The overall average type accuracy reaches 95.6\%.
These results demonstrate that AnalogToBi can generate topologies consistent with the intended circuit category.

\subsection{Case Study with SPICE Simulation}


\begin{wrapfigure}{r}{0.5\textwidth}
  \centering
  \vspace{-15pt}
  \includegraphics[width=0.42\textwidth]{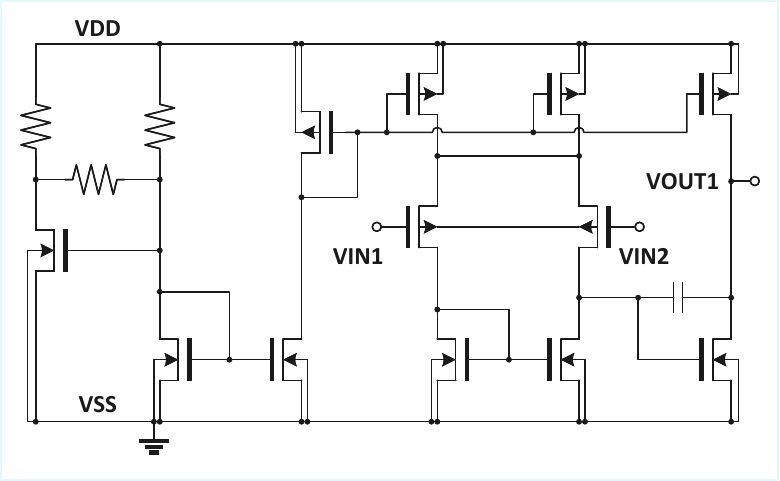}
  \includegraphics[width=0.48\textwidth]{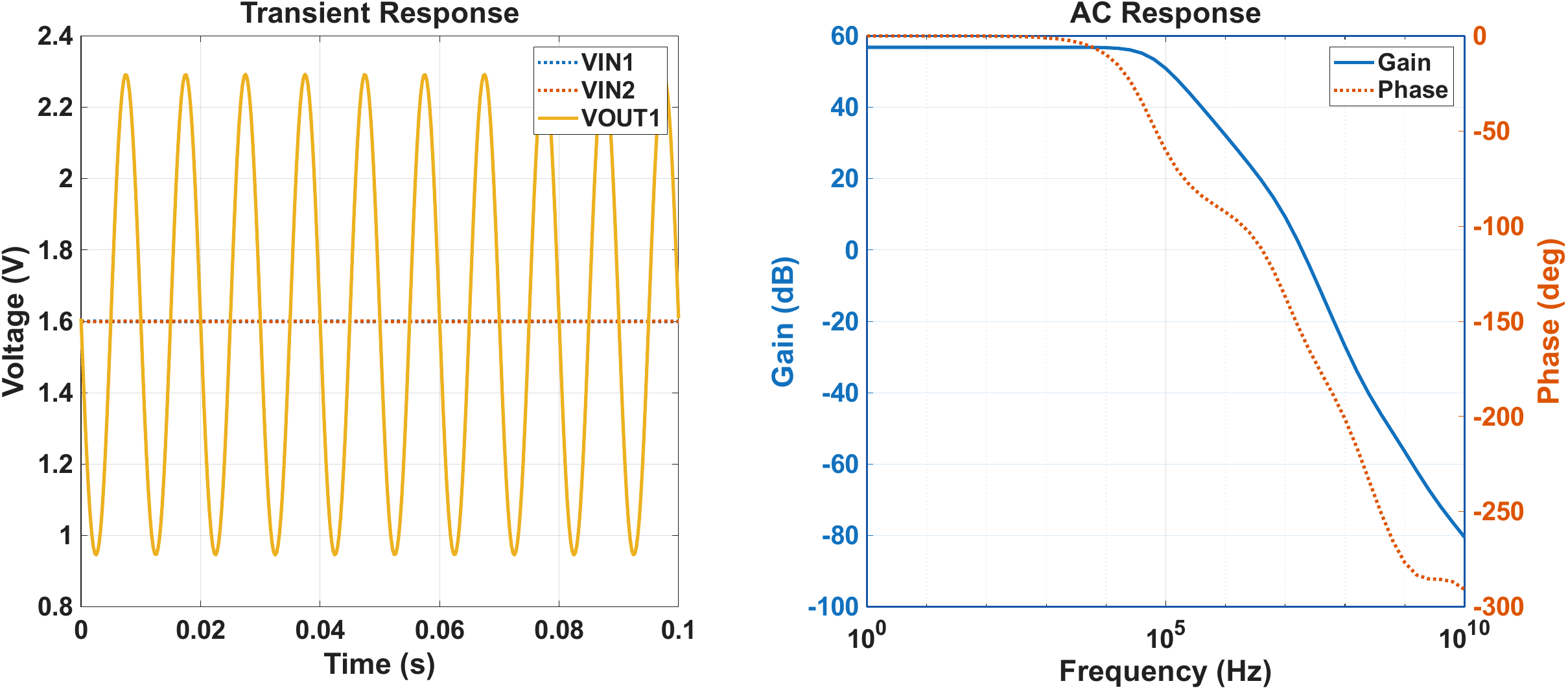}
  \caption{
  Generated OpAmp topology and transient, AC simulation result. For the transient simulation, 1~$mV_{pp}$ sinusoidal input was applied.
  }
  \label{fig:opamp_results}
  \vspace{-10pt}
\end{wrapfigure}

While the current goal of device-level topology generation is to generate electrically valid circuits without evaluating their operational behavior, we further conduct SPICE-based case studies on generated topologies.
Specifically, simulations are performed using Cadence Spectre under the GPDK 45nm CMOS process.
Following previous studies~\cite{gao2025analoggenie, gao2025analoggenielite}, we manually select heuristically promising topologies for simulation-based analysis.
For each generated topology, a default sizing procedure is applied to evaluate topology quality independently of device sizing (see Appendix~\ref{app:spice} for details).

Figure~\ref{fig:opamp_results} presents an example operational amplifier generated by AnalogToBi.
The circuit is evaluated under a 3.3V supply and a capacitive load of $100~\mathrm{pF}$ using transient and AC analyses.
The generated OpAmp achieves a FoM (MHz $\cdot$ pF/mW) of 728.1, substantially outperforming the reported result of AnalogGenie (FoM 36.5).
Even under default sizing and biasing conditions, the circuit achieves a high DC gain of 56.8~dB and moderate bandwidth characteristics (BW 58.7~kHz and GBW 40.6~MHz), exceeding the best reported GBW of AnalogGenie (12~MHz).
These results suggest that, by generating diverse and electrically valid circuits (Table~\ref{tab:main_result_1_compare_with_prev}), AnalogToBi can potentially discover practically competitive analog topologies.
Additional case studies are provided in Appendix~\ref{app:case study}.

\section{Future Work}
\label{section:future_work}

While existing device-level topology generation research primarily focuses on generating electrically valid circuits, extending these models toward properly operating circuit generation remains an important future direction.
A key limitation of prior approaches is the lack of explicit circuit-type control, which makes it difficult to define circuit-specific simulation conditions and performance objectives within a unified framework.
As a result, previous studies have largely focused on connectivity-level validity rather than operational behavior.
In contrast, AnalogToBi introduces circuit-type conditioning through circuit-type tokens, enabling generated topologies to be associated with well-defined functional categories.
This provides a natural foundation for integrating circuit-type-specific simulation pipelines into training.
For example, different simulation setups and objectives can be automatically assigned depending on the target circuit type, such as gain and bandwidth for OpAmps or switching characteristics for comparators.
Building on this capability, future work could incorporate simulation feedback directly into the optimization loop, enabling topology generation models to optimize not only electrical validity but also operational correctness and circuit performance.


\section{Conclusion}
\label{section:conclusion}
In this work, we propose AnalogToBi, a novel framework for device-level analog circuit topology generation.
By integrating circuit-type conditioning, device renaming augmentation, a bipartite graph representation, and grammar-guided decoding, AnalogToBi addresses the memorization issue and low electrical validity that limit existing approaches.
Experimental results demonstrate that AnalogToBi enables fully automated training without any HITL intervention while achieving both high validity and high novelty in generated circuit topologies.

{
\small

\bibliographystyle{unsrtnat}
\bibliography{References}

}

\clearpage
\appendix

\section{Vocabulary Definition}
\label{app:vocabulary}

\begin{table}[t]
\centering
\footnotesize
\caption{
Token categories used in AnalogToBi.
Only representative examples are shown.
}
\label{tab:vocabulary}
\begin{tabular}{l l p{0.7\linewidth}}
\toprule
Category & Example Tokens \\
\midrule
Circuit Type Token
& \texttt{CIRCUIT\_OpAmp}, \texttt{CIRCUIT\_LDO}, \texttt{CIRCUIT\_Filter}, \texttt{CIRCUIT\_Comparator} \\
\midrule
Device Token
& \texttt{NM1},\texttt{NM2}, \texttt{PM2}, \texttt{NPN1}, \texttt{R1}, \texttt{C1}, \texttt{L1}, \texttt{DIO1} \\
\midrule
Net Token
& \texttt{VSS}, \texttt{VDD}, \texttt{NET1}, \texttt{NET2}, \texttt{VIN1}, \texttt{VOUT1}, \texttt{IIN1}, \texttt{IB1} \\
\midrule
Edge (Pin) Token
& \texttt{M\_G}, \texttt{M\_S}, \texttt{M\_SB}, \texttt{M\_GD}, \texttt{B\_C}, \texttt{B\_E}, \texttt{R\_C}, \texttt{C\_C}, \texttt{L\_C} \\
\midrule
Termination Token
& \texttt{TRUNCATE} \\
\bottomrule
\end{tabular}
\end{table}

Table~\ref{tab:vocabulary} summarizes the major token categories used in AnalogToBi.
Rather than enumerating the full vocabulary, we present representative examples for clarity.

Circuit Type Tokens specify the target circuit functionality and condition the Transformer
decoder during generation.
By prepending a circuit type token (e.g., \texttt{CIRCUIT\_OpAmp} or \texttt{CIRCUIT\_LDO}),
circuit generation is formulated as a conditional sequence modeling problem.

Device Tokens represent active and passive circuit components.
These include MOSFETs, BJTs, and passive devices such as resistors, capacitors, inductors,
and diodes.
Each device instance is uniquely indexed to preserve structural identity in the generated topology.

Net Tokens correspond to electrical connection points in the circuit graph.
They represent both internal nets (e.g., \texttt{NET1}, \texttt{NET2}) and external ports
(e.g., \texttt{VSS}, \texttt{VDD}, \texttt{VIN}, \texttt{VOUT}),
enabling explicit modeling of circuit connectivity.

Edge (Pin) Tokens encode device pin--net connectivity as typed edges in the bipartite graph.
Token prefixes indicate device types
(\texttt{M\_}: MOSFET, \texttt{B\_}: BJT, \texttt{R\_}/\texttt{C\_}/\texttt{L\_}/\texttt{D\_}: resistor,
capacitor, inductor, diode),
while suffixes denote pin semantics
(e.g., \texttt{G}: gate, \texttt{S}: source, \texttt{SB}: source/body, \texttt{C}: collector).
This design allows pin semantics to be shared across devices while keeping the graph representation compact.

Finally, the \texttt{TRUNCATE} token is used for sequence padding after generation terminates,
indicating that no further structural tokens follow.

\begin{figure}[h!]
    \centering
    \includegraphics[width=\linewidth]{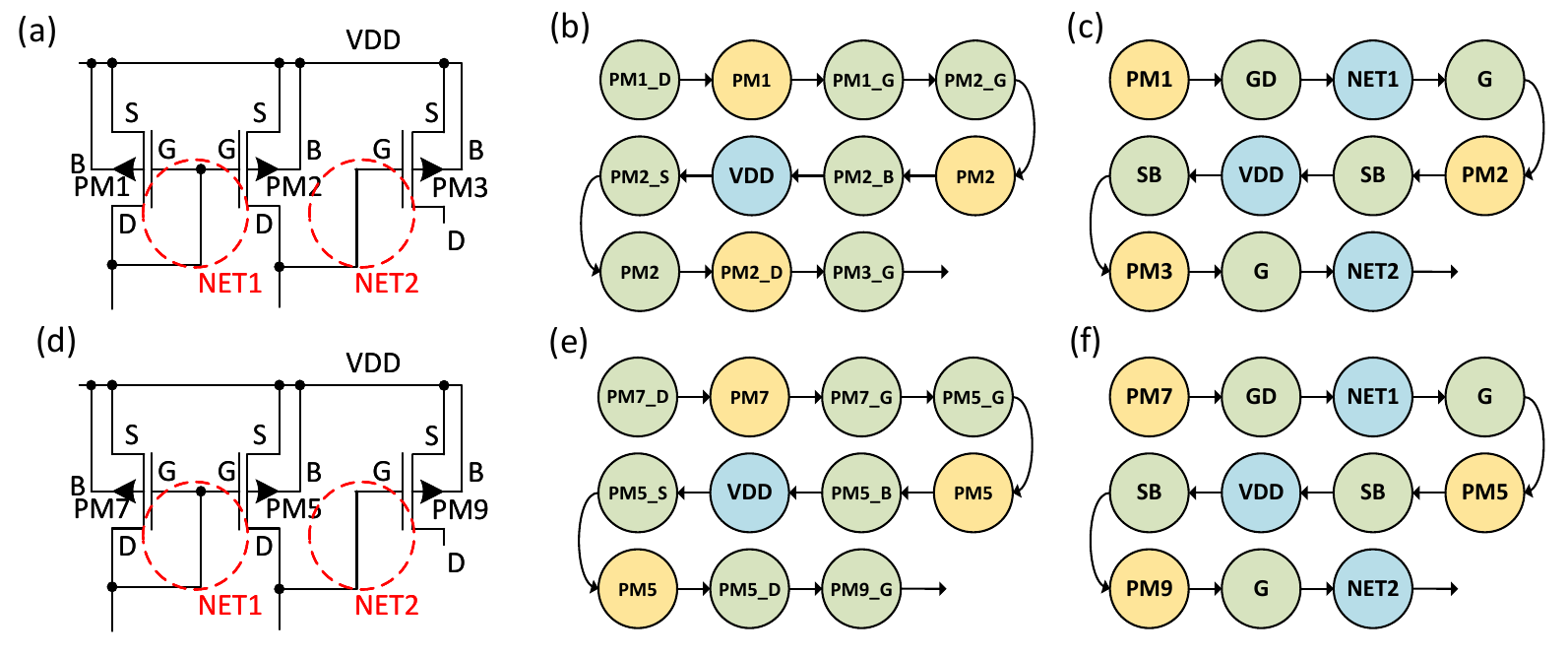}
    \caption{Examples of device renaming augmentation applied to circuit graphs serialized into token sequences.
    (a), (d) show the original and renamed circuit graphs.
    (b), (e) show the renamed sequences under the device--pin representation, and (c), (f) show the renamed sequences under the bipartite representation.
    In both representations, only the token index assigned to each device are randomly shuffled while the underlying structural connectivity remains unchanged.}
    \label{fig:renaming}
\end{figure}

\section{Examples of Renaming Augmentation}
\label{app:renaming}

Fig.~\ref{fig:renaming} illustrates examples of the renaming augmentation applied to circuit graphs serialized into token sequences.
Device renaming augmentation can be applied to both the device--pin representation (b), (e) and the bipartite representation (c), (f).
In both cases, only the token indices assigned to each device are randomly shuffled, while the underlying structural connectivity of the circuit remains unchanged, as shown in the original circuits (a), (d).
As such, renaming augmentation generates structurally equivalent but tokenically distinct sequences, encouraging the model to focus on learning circuit topology rather than memorizing specific token patterns.

\section{Model Architecture and Training Details}
\label{app:model}

\subsection{GPT Decoder}
\label{app:gpt}

We use a decoder-only Transformer language model for autoregressive circuit generation,
following the standard GPT-style architecture \cite{vaswani2017attention, radford2019gpt2}.

The model uses an embedding dimension of 384 with 6 attention heads and 6 Transformer layers.
The maximum context length is fixed to 1024 tokens, and a dropout rate of 0.2 is applied throughout
the network.
All linear and embedding layers are initialized from $\mathcal{N}(0, 0.02)$.
The resulting model contains approximately 11.3M parameters.

Training is performed using AdamW with a learning rate of $3\times10^{-4}$ and batch size 64
for 100{,}000 iterations.
Validation is conducted every 500 iterations, and the model checkpoint achieving the lowest validation loss (0.2763 at step 97,500) is selected; the entire training process takes 22.8 hours on an NVIDIA RTX 5880.

During inference, circuits are generated autoregressively using multinomial sampling with a fixed
temperature of 0.7 and a maximum sequence length of 1024.
Grammar-guided decoding (Section~\ref{sec:grammar_guided_decoding}) is applied as an external constraint by masking invalid token
transitions.
Under this setting, the average inference latency is 1.45 s per sample.

\subsection{GAT Classifier}
\label{app:gat}

We train a Graph Attention Network (GAT) classifier \cite{velickovic2018gat} to evaluate whether
generated circuit topologies match the intended circuit type.
The classifier is used exclusively for evaluation and does not provide feedback to the generator.

Node and edge(pin)-type tokens are embedded with dimensionality 64, and edge embeddings are linearly projected to match the edge feature dimension used in attention. The classifier consists of three stacked GAT layers, using four attention heads in the first two layers and a single attention head in the final layer. Batch normalization is applied after each GAT layer, while ELU activation and dropout with rate 0.3 are used after all but the final layer. The resulting model contains 1.08M parameters.

During training and inference, each circuit is represented as a graph $G=(V,E)$, where nodes correspond to devices or nets and
edges encode connectivity types (i.e., pins) between them.

For the $k$-th attention head, the attention coefficient between nodes $i$ and $j$ is computed as
\begin{equation}
e_{ij}^{(k)} =
\mathrm{LeakyReLU}\!\left(
\mathbf{a}_k^\top
\big[ \mathbf{W}_k \mathbf{h}_i \,\Vert\, \mathbf{W}_k \mathbf{h}_j \,\Vert\, \mathbf{e}_{ij} \big]
\right),
\end{equation}
and the normalized attention coefficient is given by
\begin{equation}
\alpha_{ij}^{(k)} = \mathrm{softmax}_{j}\!\left( e_{ij}^{(k)} \right),
\end{equation}
where $\mathbf{h}_i$ and $\mathbf{h}_j$ denote node embeddings and $\mathbf{e}_{ij}$ represents
the corresponding edge-type embedding.

A permutation-invariant pooling operation is applied to obtain a graph-level representation
\begin{equation}
\mathbf{h}_G =
\mathrm{Pool}
\left(
\left\{
\mathbf{h}_i^{(L)} \mid i \in V
\right\}
\right),
\end{equation}
which serves as a graph-level embedding capturing the global circuit topology.
This embedding is followed by a two-layer multilayer perceptron to predict one of 15 circuit
categories.

The classifier is trained for 100 epochs using the Adam optimizer with learning rate \(1\times10^{-4}\) and weight decay \(10^{-3}\), with dataset ratio weighted cross-entropy loss handling rare class dataset. A cosine annealing learning-rate schedule and gradient clipping with maximum norm 1.0 are applied to stabilize training. The model is trained on an NVIDIA RTX 5880 GPU for 1.86 hours, achieving a Top-1 accuracy of 98.61\% on the training set and 91.33\% on validation set, with an average inference latency of 2.02 ms per sample.

\begin{figure}[t]
    \centering
    \vspace{-4pt}
    \includegraphics[width=\linewidth]{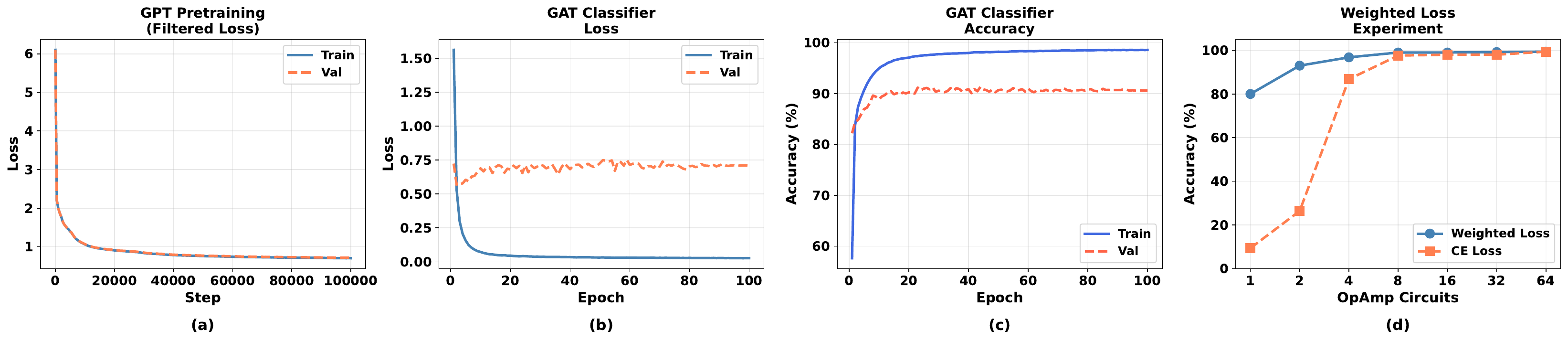}
    \caption{(a) is training curve of GPT model, (b) and (c) are training curve and accuracy of GAT classifier showing lower bound of validation result due to per circuit ID split, (d) shows effectiveness of dataset ratio weighted CE loss result versus conventional CE loss of GAT classifier}
    \label{fig:training_curves}
    \vspace{-4pt}
\end{figure}

\subsection{Training Curves and Accuracy}
\label{app:training_curves}

Figure~\ref{fig:training_curves} report the training
dynamics of the (a) GPT decoder and the (b),(c) GAT classifier, respectively. Figure (d) illustrates the effect of class-weighted loss employed in our GAT-based circuit type classifier. Since our training dataset is inherently imbalanced due to the presence of rare circuit classes, we adopt class-weighted cross-entropy loss, which inversely scales the loss contribution of each class according to its frequency in the training set. 

For a controlled evaluation of class imbalance, we construct the dataset as follows. We include 100 samples each from three non-target circuit classes—BGR, LDO, and Power Converter—and split them into training and validation sets with a 9:1 ratio. In addition, we vary the number of OpAmp circuits included in the training set (ranging from 1 to 64) to simulate different levels of data scarcity for the target class. The validation set remains fixed across all settings.

All experiments are conducted using the same GAT architecture and hyperparameter configuration described in the main text, ensuring that performance differences arise solely from the loss formulation. Classification accuracy is evaluated using a held-out set of 500 unseen OpAmp topologies that are not used during GAT training, ensuring a fair assessment of generalization performance.

As shown in the results, class-weighted loss consistently outperforms standard cross-entropy across all data scales, most notably in the extreme few-shot regime where it achieves 80.00\% and 93.00\% accuracy with only one and two OpAmp training samples, compared to 9.40\% and 26.40\% for the non-weighted baseline, demonstrating its effectiveness in handling class imbalance under scarce supervision.

\section{Grammar-Guided Decoding Specification}
\label{app:grammar}

The grammar-guided decoding mechanism constrains autoregressive generation through rule-based
token constraint. This appendix formalizes the induced states, transition
rules, and termination conditions.

\subsection{State Definitions and Decoding Buffers}
\label{app:grammar_state}

Decoding states are defined by the semantic roles of the most recent tokens.
Let the tokens at time steps $T-1$ and $T$ denote the two most recently generated tokens.
The ordered token pair $(t_{T-1}, t_T)$ uniquely determines the current decoding state
and the set of admissible next tokens.

We define the following six states:

\begin{table}[t]
\centering
\footnotesize
\caption{State definitions used in the grammar guided decoding process.}
\label{tab:states}

\begin{tabular}{p{0.23\linewidth} p{0.70\linewidth}}
\toprule
\textbf{State} & \textbf{Description} \\
\midrule
\midrule

\textbf{Circuit Type -- VSS} 
& Initial decoding state, defined by a circuit-type token followed by the reference ground \texttt{VSS}. \\

\midrule

\textbf{Net -- Edge} 
& A state in which a net or external port token is followed by an edge-type token. \\

\midrule

\textbf{Edge -- Device} 
& A state in which an edge-type token is followed by a device token. \\

\midrule

\textbf{Device -- Edge} 
& A state in which a device token is followed by an edge-type token. \\

\midrule

\textbf{Edge -- Net} 
& A state in which an edge-type token is followed by a net or external port token. \\

\midrule

\textbf{Edge -- VSS} 
& A special case of Edge--Net in which the net token corresponds to the reference ground \texttt{VSS}. \\

\bottomrule
\end{tabular}
\end{table}

Transitions between decoding states are enforced through token masking at each time step.
At time $T+1$, the set of admissible tokens is determined by the semantic types of the tokens
generated at time steps $T-1$ and $T$, together with auxiliary decoding buffers that track
partial connectivity.

\begin{table}[t]
\centering
\footnotesize
\setlength{\tabcolsep}{4pt}
\caption{Decoding buffers in grammar guided decoding}
\label{tab:buffers}

\begin{tabular}{p{0.19\textwidth}  p{0.27\textwidth} p{0.47\textwidth}}
\toprule
\textbf{Buffers} & \textbf{Tracks} & \textbf{Role}\\
\midrule
\midrule

\texttt{device-pins}
& Which pins of each device have been connected.
& All required pins must be used before TRUNCATE is allowed (MOSFET: D/G/S/B; BJT: B/C/E; Diode: P/N). \\

\midrule

\texttt{net\_connections}
& Which devices are connected to each internal net
& Every internal net must connect to $\geq2$ distinct devices (floating net prevention). \\

\midrule

\texttt{internal\_nets\_seen}
& Internal nets (NET1–NET50) that have appeared in the sequence.
& Verify connectivity for all nets that have been generated. \\

\midrule

\texttt{device\_pins}
& Which pins of each device have been connected.
& All required pins must be used before TRUNCATE is allowed (MOSFET: D/G/S/B; BJT: B/C/E; Diode: P/N) \\

\midrule

\texttt{device\_edge\_nets}
& Which net a specific (device, edge) pair is already assigned to.
& Prevents the same edge from connecting to two different nets; allows reconnection to the same net \\

\midrule

\texttt{passive\_net\_count}
& Nets connected to each 2-terminal device (R/C/L or Diode).
& Passives connect to exactly 2 distinct nets; diode edges must connect to distinct nets. \\

\midrule

\texttt{device\_pin\_nets}
& Which nets each individual pin of an active device is assigned to.
& Prevents pin-level conflicts for multi-pin edges (e.g., \texttt{M\_BD} covers both B and D pins simultaneously). \\

\bottomrule
\end{tabular}
\end{table}

\paragraph{Decoding buffers.}
Decoding buffers and its role in grammar guided decoding is explained in table~\ref{tab:buffers}. All buffers are updated incrementally at each decoding step ($O(1)$ per token), avoiding full sequence rescans enabling efficient batch generation.

\subsection{Transition Rules}
\label{app:grammar_transition}


\paragraph{State transitions.}

\begin{itemize}
    \item \textbf{Circuit Type -- VSS $\rightarrow$ Net -- Edge.} \\
    This transition corresponds to the case where the token at time $T-1$ is a circuit-type token
    and the token at time $T$ is the reference ground net \texttt{VSS}.
    At time $T+1$, the decoder permits edge-type tokens to initiate the traversal of the bipartite
    graph from the reference node.

    \item \textbf{Net -- Edge $\rightarrow$ Edge -- Device.} \\
    This transition corresponds to the case where the token at time $T-1$ is a net or external port
    (e.g., \texttt{VSS}, \texttt{VIN}, \texttt{NET*}) and the token at time $T$ is an edge-type token
    (e.g., \texttt{M\_S}, \texttt{M\_D}).
    Under this state, the token at time $T+1$ must be a device token compatible with the semantics
    of the edge type.

    \item \textbf{Edge -- Device $\rightarrow$ Device -- Edge.} \\
    This transition corresponds to the case where the token at time $T-1$ is an edge-type token and
    the token at time $T$ is a device token.
    At time $T+1$, the decoder permits edge-type tokens associated with the device category of
    $t_T$, enforcing type-level consistency.

    \item \textbf{Device -- Edge $\rightarrow$ Edge -- Net.} \\
    This transition corresponds to the case where the token at time $T-1$ is a device token and the
    token at time $T$ is an edge-type token.
    At time $T+1$, the decoder restricts the admissible tokens to net or external port tokens, while
    recording the resulting device--edge--net association.

    \item \textbf{Edge -- Net $\rightarrow$ Net -- Edge.} \\
    This transition corresponds to the case where the token at time $T-1$ is an edge-type token and
    the token at time $T$ is a net or external port token.
    Decoding proceeds by generating another edge-type token incident to the current net.

    \item \textbf{Edge -- Net $\rightarrow$ Edge -- VSS.} \\
    This transition is a special case of Edge--Net where the token at time $T$ is \texttt{VSS}.
    At time $T+1$, the decoder permits edge-type tokens as usual; additionally, the \texttt{TRUNCATE}
    token becomes admissible only if the electrical rule checks (e.g., complete pin usage and internal-net
    connectivity) are satisfied.
\end{itemize}

At each decoding step, tokens that violate the above transition rules are masked out and assigned
zero probability.

\subsection{Termination Conditions}
\label{app:grammar_termination}

Decoding terminates when one of the following conditions is satisfied:
\begin{enumerate}
  \item A \texttt{TRUNCATE} token is generated in the Edge--Net state with \texttt{VSS} at time $T$,
  and structural constraints are satisfied, including complete assignment of device terminals
  and valid connectivity of all internal nets.
  \item The maximum allowed sequence length is reached.
  \item No valid tokens remain after masking, indicating that decoding has reached a dead-end under
the grammar constraints (e.g., due to terminal usage or reconnection restrictions), and the
sequence is marked as invalid.
\end{enumerate}

These termination conditions ensure that decoding halts only after the bipartite traversal
returns to the reference ground node \texttt{VSS} and essential structural requirements are fulfilled.

\clearpage

\begin{algorithm}[t]
  \caption{Structure-Preserving Data Augmentation for Bipartite Circuit Sequences}
  \label{alg:augmentation}
  \begin{algorithmic}
    \STATE {\bfseries Input:} Bipartite circuit graph $G=(V,E)$ with typed edges; reference node $v_{\mathrm{ref}}=\texttt{VSS}$;
    maximum augmented sequences $K$; maximum sequence length $T_{\max}=1024$.
    \STATE {\bfseries Output:} Augmented sequence set $\mathcal{S}$.

    \STATE Initialize $\mathcal{S} \leftarrow \emptyset$.
    \FOR{$k=1$ {\bfseries to} $K$}
      \STATE Initialize $s \leftarrow [\,]$, $v \leftarrow v_{\mathrm{ref}}$.
      \STATE Initialize visited edge set $\mathcal{E}_{\mathrm{vis}}\leftarrow \emptyset$ and visited node set $\mathcal{V}_{\mathrm{vis}}\leftarrow \{v_{\mathrm{ref}}\}$.
      \STATE Append $v$ to $s$.
      \STATE Randomly shuffle neighbor orderings for all nodes in $G$.
      
      \WHILE{$\mathcal{E}_{\mathrm{vis}} \neq E$}
        \STATE Choose next step using the following priority:
        \STATE \hspace{0.8em}(i) an unvisited incident edge $e=(v,u)$;
        \STATE \hspace{0.8em}(ii) an edge leading to an unvisited node $u$;
        \STATE \hspace{0.8em}(iii) a shortest path from $v$ to any node incident to an unvisited edge.
        \STATE Let the chosen move be $v \rightarrow e \rightarrow u$ (allow revisits if needed for coverage).
        \STATE Append edge-type token $\tau(e)$ and node token $u$ to $s$.
        \STATE Update $\mathcal{E}_{\mathrm{vis}}\leftarrow \mathcal{E}_{\mathrm{vis}} \cup \{e\}$ and $\mathcal{V}_{\mathrm{vis}}\leftarrow \mathcal{V}_{\mathrm{vis}} \cup \{u\}$.
        \STATE Set $v \leftarrow u$.
      \ENDWHILE

      \STATE \textbf{Close the loop:} append a path from current $v$ back to $v_{\mathrm{ref}}$ (if $v \neq v_{\mathrm{ref}}$), appending corresponding $\tau(e)$ and node tokens along the path.
      \STATE \textbf{Coverage check:} if $\mathcal{V}_{\mathrm{vis}} \neq V$, discard this sequence and {\bfseries continue}.
      \STATE \textbf{ERC check:} if $\mathrm{ERC}(s)$ fails (e.g., floating internal nets or incomplete device connections), discard this sequence and {\bfseries continue}.
            
      \STATE Pad $s$ with \texttt{TRUNCATE} to obtain fixed length $T_{\max}$.
      \STATE Add $s$ to $\mathcal{S}$.
    \ENDFOR

    \STATE \textbf{return} $\mathcal{S}$.
  \end{algorithmic}
\end{algorithm}

\section{Data Augmentation Algorithm}
\label{app:augmentation}

We summarize the structure-preserving data augmentation procedure used in this work
in Algorithm~\ref{alg:augmentation}.
The algorithm generates multiple distinct token sequences from a single bipartite
circuit graph by varying traversal orders while preserving exact topological structure.
All augmented sequences are required to cover every device and net, satisfy electrical
rule checks, and remain compatible with grammar-guided decoding.

Here, $G=(V,E)$ denotes a bipartite circuit graph with device and net nodes,
and $\tau(e)$ denotes the discrete edge-type token associated with edge $e$,
encoding pin or connection semantics.
$\mathcal{V}_{\mathrm{vis}}$ and $\mathcal{E}_{\mathrm{vis}}$ represent the sets
of visited nodes and edges during sequence construction, respectively.
$K$ controls the maximum number of augmented sequences generated per circuit,
and $T_{\max}$ denotes the fixed maximum sequence length after padding.
$\mathrm{ERC}(s)$ denotes an electrical rule check that filters out sequences
with invalid connectivity, such as floating internal nets or incomplete device connections.

\begin{figure}[t]
\centering

\begin{subfigure}{0.30\linewidth}
    \centering
    \includegraphics[width=\linewidth]{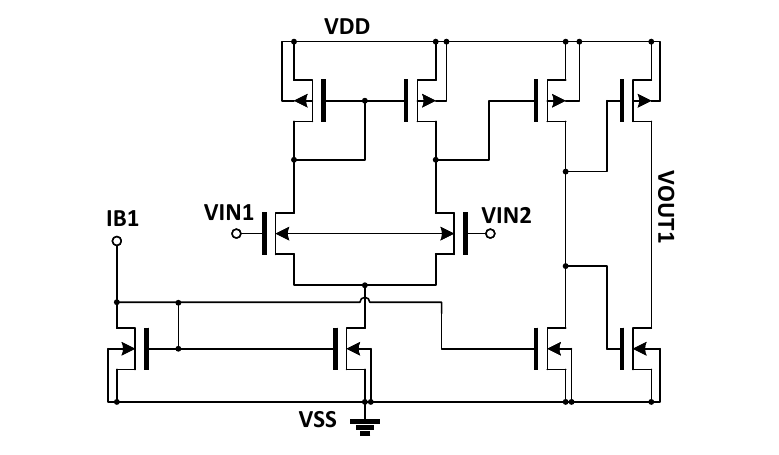}
    \caption{Generated comparator}
\end{subfigure}
\hfill
\begin{subfigure}{0.30\linewidth}
    \centering
    \includegraphics[width=\linewidth]{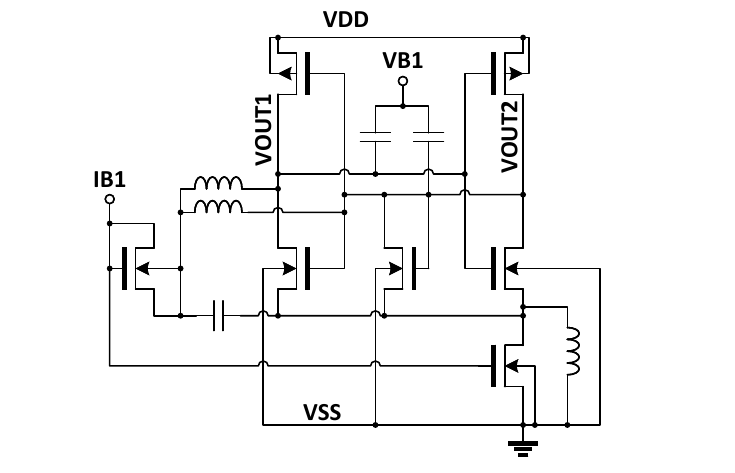}
    \caption{Generated oscillator topology}
\end{subfigure}
\hfill
\begin{subfigure}{0.30\linewidth}
    \centering
    \includegraphics[width=\linewidth]{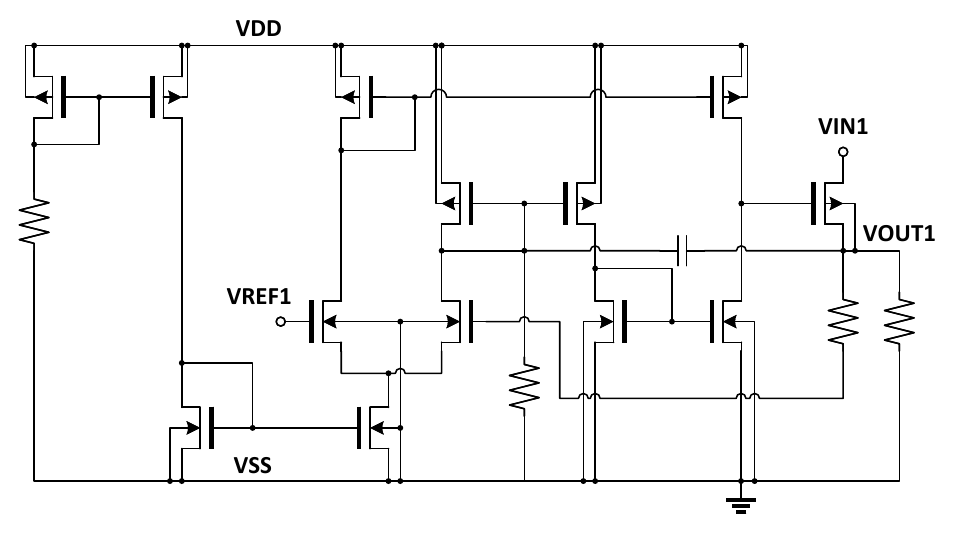}
    \caption{Generated LDO topology}
\end{subfigure}


\begin{subfigure}{0.30\linewidth}
    \centering
    \includegraphics[width=\linewidth]{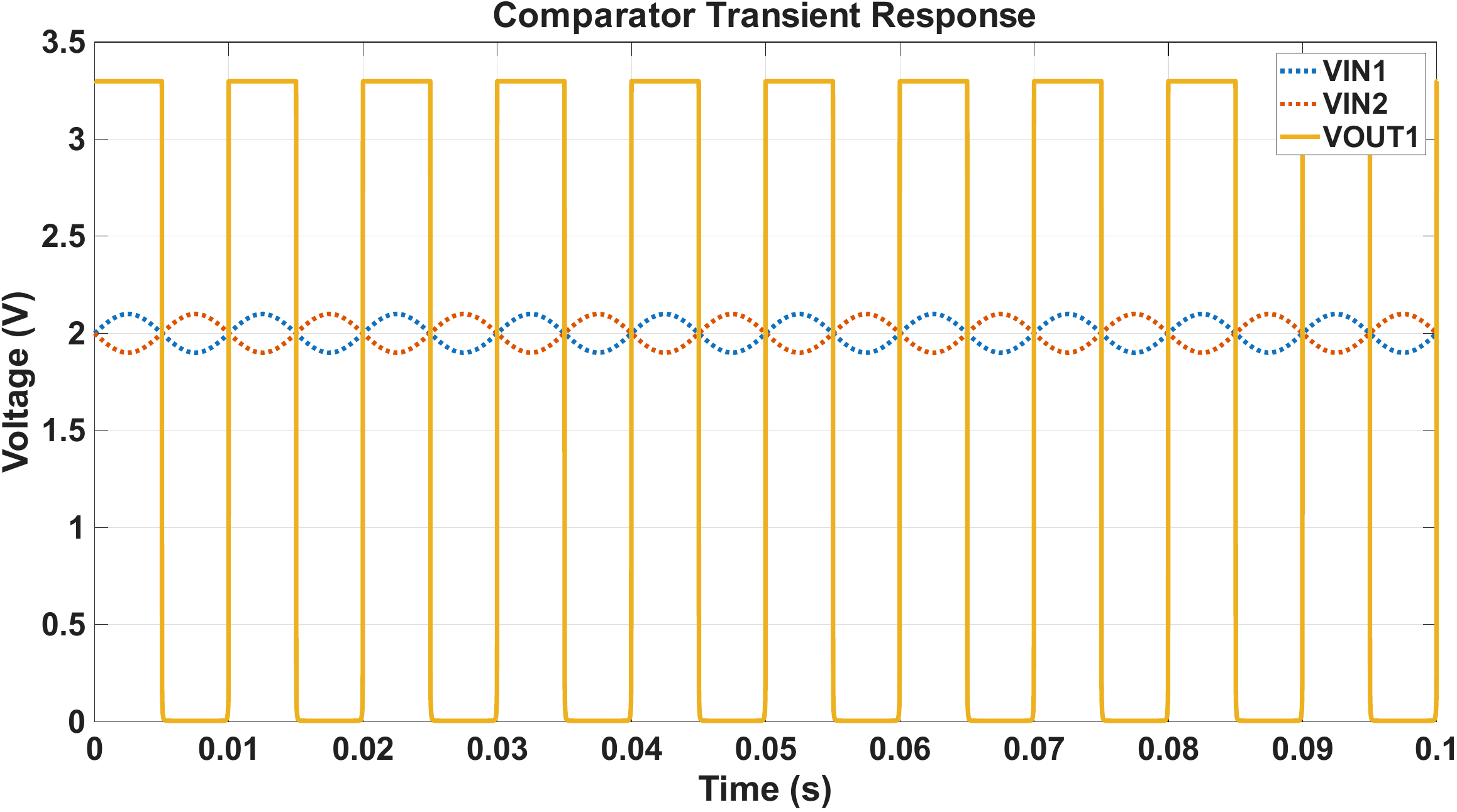}
    \caption{Comparator transient response}
\end{subfigure}
\hfill
\begin{subfigure}{0.30\linewidth}
    \centering
    \includegraphics[width=\linewidth]{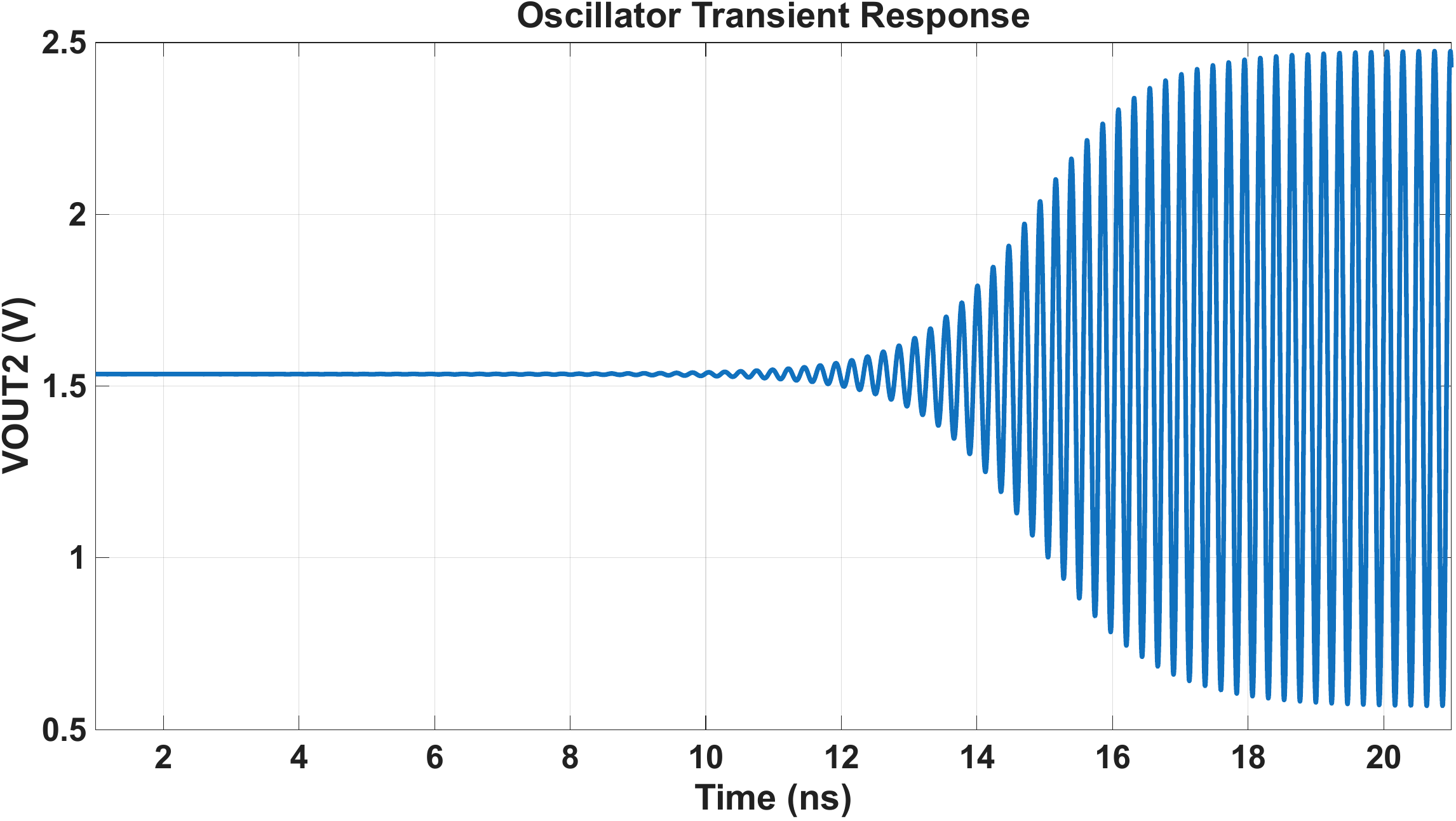}
    \caption{Oscillator transient response}
\end{subfigure}
\hfill
\begin{subfigure}{0.30\linewidth}
    \centering
    \includegraphics[width=\linewidth]{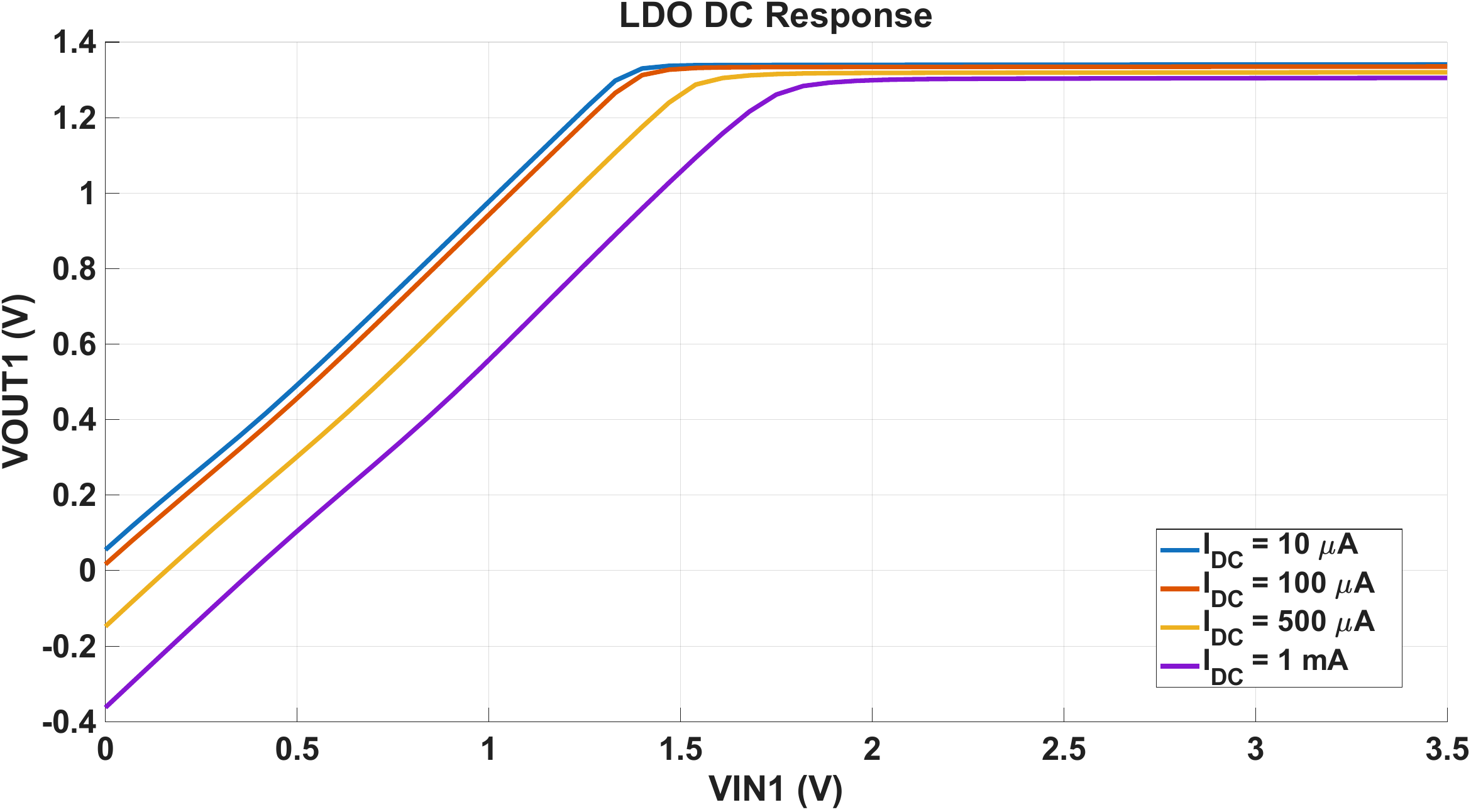}
    \caption{LDO DC simulation result across current load}
\end{subfigure}

\caption{
Comparator, oscillator and LDO topologies with their corresponding transient/DC simulation result.
}
\label{fig:combined_results}
\end{figure}

\section{Sequence-to-SPICE Translation with default sizing}
\label{app:spice}

The generated circuits were translated into netlists with default biasing and sizing, enabling simulation in Cadence Spectre. 

For biasing, a supply voltage of VDD = 3.3 V is used. Gate bias voltages are assigned based on device type: 0.75·VDD for PMOS-connected nodes and 0.25·VDD for NMOS-connected nodes. If unspecified, a default bias of 0.5·VDD is applied. For current biasing, a nominal current of 100 $\mu$A is used, with current direction defined as entering NMOS drains and leaving PMOS drains.

For device sizing, NMOS transistors are assigned W/L = 10$\mu$m/1$\mu$m, PMOS transistors W/L = 15$\mu$m/1$\mu$m, resistors R = 10 k$\Omega$, capacitors C = 1 pF, and inductors L = 1 nH.

\section{Additional Case Study}
\label{app:case study}
\subsection{Comparator Case Study with Transient Simulation}

This section covers several more samples of the topology created by AnalogToBi.
Figure~\ref{fig:combined_results} (a) shows the generated transistor-level comparator circuit.
The topology comprises a differential input stage, a pre-amplification path,
and a pull-up/pull-down output stage, forming a structure consistent with dynamic comparator
architectures.

To evaluate its behavior, the generated sequence is automatically translated into a SPICE
netlist using the sequence-to-SPICE decoder with default sizing and biasing logic and simulated under a 45nm CMOS process under Cadence Spectre.
The comparator is driven by a $100\,\mathrm{mV}$ peak differential sinusoidal input centered around the common-mode voltage.

Figure~\ref{fig:combined_results} (d) presents the corresponding transient response obtained from 
simulation.
The output exhibits clear comparator behavior and switches decisively between the supply rails.
Specifically, the output node transitions fully from 3.3V VDD to VSS, demonstrating
robust rail-to-rail swing and confirming correct comparator functionality.

\subsection{Oscillator Case Study with Transient Simulation}

Figures~\ref{fig:combined_results} (b) illustrates the generated oscillator topology, which exhibits a cross-coupled structure with an LC tank, consistent with LC oscillator designs. The topology includes differential output nodes (VOUT1 and VOUT2), a biasing network (VB1), and an inductive load, forming a feedback loop that sustains oscillation.

Figures~\ref{fig:combined_results} (e) shows the corresponding transient response obtained from SPICE simulation. The output initially remains near the bias point and then gradually builds up oscillation amplitude due to the inherent startup mechanism driven by noise or small perturbations. As time progresses, the amplitude increases exponentially until it reaches a steady-state limit cycle, indicating that the loop gain exceeds unity during startup and is eventually stabilized by nonlinear effects. The measured oscillation frequency is 4.27 GHz.

\subsection{LDO Case Study with DC Simulation}

Figures~\ref{fig:combined_results} (c) illustrates an LDO topology generated by AnalogToBi, consisting of a pass transistor, an error amplifier, and internal biasing circuitry. The output voltage (VOUT1) is fed back through a resistive divider to the error amplifier, where it is compared against the reference voltage (VREF1). The error signal, the amplified gap between the feedback voltage and VREF1, controls the pass device gate, forming a negative feedback loop that regulates VOUT1 against variations in input voltage (VIN1) and load current.

Figures~\ref{fig:combined_results} (f) demonstrates stable output voltage across varying load currents and input voltages. The LDO achieves a load regulation of 10.8 mV/mA and a line regulation of 0.51 mV/V, indicating robust regulation performance. Considering the current-driving capability of the default-sized pass transistor, we perform the simulation with load current sweep to 1 mA.

\section{Detailed Ablation Study Results by Circuit Type}

\begin{table}
\centering
\caption{%
Detailed ablation study results by Circuit Type. These results are obtained under a setting that enables per-type metric evaluation via circuit-type tokens. For each circuit type, 1,000 inference samples are generated. %
}
\label{tab:ablation_detail}

\resizebox{\linewidth}{!}{
\begin{tabular}{lccccccccccccccc|c}
\toprule
\multicolumn{17}{c}{\textbf{Device-pin Representation + Type Token}} \\
\midrule
 & OpAmp & Mirror & Comp & Mix & LDO & Oscil & Filt & BGR & PAmp & VoltR & PConv & PLL & S Cap & A/D\_D/A & General & Avg \\

\midrule
\multicolumn{17}{l}{\textbf{-- 1000 samples per circuit type}} \\

Validity (\%)
& 88.0&	93.4&	79.3&	84.9&	68.4&	84.8&	84.0&	91.2&	86.2&	90.2&	95.4&	87.8&	85.3&	84.1&	89.1&	86.1 \\

Novelty (\%)
&30.4	&25.3	&35.7	&31.1	&54.0	&30.0	&29.3	&24.2	&28.3	&20.3	&12.5	&29.8	&31.6	&25.0	&30.4	&29.2\\

Valid \& Novel (\%) 
&18.4	&18.7	&15.1	&16.0	&22.4	&14.8	&13.4	&15.5	&14.5	&10.5	&7.9	&17.6	&16.9	&9.6	&19.5	&15.4\\

Exact. (\%)
&56.7	&58.6	&53.9	&53.4	&37.4	&57.6	&56.0	&61.7	&59.4	&63.8	&72.8	&56.3	&60.3	&59.0	&54.2	&57.5\\

Type Acc. (\%) 
&96.3	&89.4	&94.3	&99.5	&99.7	&95.3	&83.2	&97.6	&96.3	&90.7	&99.7	&93.2	&99.2	&90.2	&--	&94.6\\
															
\midrule
\midrule
\multicolumn{17}{c}{\textbf{Device-pin Representation + Type Token + Renaming}} \\
\midrule
 & OpAmp & Mirror & Comp & Mix & LDO & Oscil & Filt & BGR & PAmp & VoltR & PConv & PLL & S Cap & A/D\_D/A & General & Avg \\
\midrule
Validity (\%)      
&71.5	&85.0	&37.9	&69.0	&30.8	&61.8	&45.2	&55.6	&60.2	&46.6	&37.8	&63.6	&56.5	&42.5	&69.8	&55.6\\

Novelty (\%)        
&75.4	&67.3	&90.7	&86.6	&90.9	&81.0	&94.2	&75.7	&93.0	&87.8	&88.8	&82.2	&75.4	&90.2	&82.4	&84.1\\

Valid \& Novel (\%) 
&46.9	&52.3	&28.6	&55.6	&21.7	&42.8	&39.4	&31.3	&53.2	&34.4	&26.6	&45.8	&31.9	&32.7	&52.2	&39.7\\

Exact. (\%)       		
&0.0	&0.0	&0.0	&0.0	&0.0	&0.0	&0.0	&0.0	&0.0	&0.0	&0.0	&0.0	&0.0	&0.0	&0.0	&0.0\\

Type Acc. (\%) 
&90.1	&85.8	&92.5	&98.9	&92.2	&88.6	&94.8	&93.5	&87.9	&73.6	&90.4	&86.3	&95.2	&82.3	&--	&89.4\\
															
\midrule
\midrule
\multicolumn{17}{c}{\textbf{Bipartite representation + Type Token}} \\
\midrule
 & OpAmp & Mirror & Comp & Mix & LDO & Oscil & Filt & BGR & PAmp & VoltR & PConv & PLL & S Cap & A/D\_D/A & General & Avg \\
\midrule
Validity (\%)       
&64.2	&65.0	&45.4	&76.0	&27.7	&49.8	&56.4	&66.1	&52.2	&63.2	&36.6	&45.7	&45.8	&80.1	&59.2	&55.6\\

Novelty (\%)        
&48.0	&51.1	&62.9	&46.9	&81.9	&68.8	&50.9	&49.5	&65.4	&43.2	&72.8	&65.9	&58.2	&34.2	&59.5	&57.3\\

Valid \& Novel (\%) 
&13.1	&16.6	&9.2	&23.1	&10.0	&19.1	&8.2	&16.3	&18.2	&7.3	&10.4	&12.9	&5.3	&14.7	&19.6	&13.6\\

Exact. (\%)       
&1.5	&11.4	&0.3	&6.0	&0.0	&6.5	&3.0	&0.0	&5.7	&12.2	&0.0	&3.9	&0.0	&0.0	&6.2	&3.8\\

Type Acc. (\%) 
&97.8	&89.5	&88.2	&99.9	&100.0	&98.0	&98.7	&100.0	&95.5	&99.2	&100.0	&95.3	&98.6	&99.9	&--	&97.2\\
															
\midrule
\midrule




															
\multicolumn{17}{c}{\textbf{Bipartite representation + Type Token + Grammar}} \\
\midrule
 & OpAmp & Mirror & Comp & Mix & LDO & Oscil & Filt & BGR & PAmp & VoltR & PConv & PLL & S Cap & A/D\_D/A & General & Avg \\
\midrule
Validity (\%)       
&98.9	&99.5	&97.4	&99.8	&97.3	&95.8	&99.7	&99.1	&93.5	&99.4	&96.7	&97.2	&97.8	&99.4	&96.0	&97.8\\

Novelty (\%)        
&78.8	&69.3	&71.6	&72.9	&92.3	&83.4	&38.6	&88.0	&87.9	&69.6	&98.5	&77.8	&68.4	&44.7	&77.3	&74.6\\

Valid \& Novel (\%) 
&77.7	&68.8	&69.0	&72.7	&89.6	&79.2	&38.3	&87.1	&81.4	&69.0	&95.2	&75.0	&66.2	&44.1	&73.3	&72.4\\

Exact. (\%)       
&1.6	&3.5	&0.0	&1.4	&0.0	&4.0	&2.8	&0.0	&0.8	&2.5	&0.0	&2.5	&0.0	&0.0	&2.3	&1.4\\

Type Acc. (\%) 
&95.7	&83.8	&87.4	&99.9	&99.3	&97.4	&97.8	&99.9	&91.0	&96.6	&100.0	&88.3	&92.4	&99.3	&--	&94.9\\
															
\midrule
\midrule
\multicolumn{17}{c}{\textbf{Bipartite representation + Type Token + Renaming + Grammar (Proposed)}} \\
\midrule
 & OpAmp & Mirror & Comp & Mix & LDO & Oscil & Filt & BGR & PAmp & VoltR & PConv & PLL & S Cap & A/D\_D/A & General & Avg \\
\midrule
Validity (\%)       
& 97.6 & 99.4 & 97.1 & 99.3 & 95.6 & 97.7 & 98.9 & 97.7 & 97.7 & 98.5 & 96.2 & 97.8 & 95.9 & 99.6 & 98.1 & \textbf{97.8} \\

Novelty (\%)        
& 89.1 & 77.4 & 92.0 & 84.1 & 97.6 & 92.1 & 92.9 & 94.0 & 99.7 & 91.0 & 99.2 & 90.9 & 91.4 & 96.5 & 92.9 & \textbf{92.1} \\

Valid \& Novel (\%) 
& 86.7 & 76.8 & 89.1 & 83.5 & 93.2 & 89.8 & 91.8 & 91.7 & 97.4 & 89.5 & 95.4 & 88.7 & 87.3 & 96.1 & 91.0 & \textbf{89.9} \\

Exact. (\%)       
&0.0	&0.0	&0.0	&0.0	&0.0	&0.0	&0.0	&0.0	&0.0	&0.0	&0.0	&0.0	&0.0	&0.0	&0.0	&\textbf{0.0}\\															

Type Acc. (\%) 
&95.8	&92.4	&86.4	&99.6	&100.0	&96.7	&93.6	&100.0	&95.3	&98.5	&100.0	&89.6	&95.2	&95.1	&--	&\textbf{95.6}\\

\bottomrule
\end{tabular}
}
\end{table}

Table~\ref{tab:ablation_detail} reports detailed ablation results broken down by circuit type, complementing the
aggregate analysis presented in Section~\ref{sec:ablation}.
Across nearly all circuit categories, the bipartite representation combined with grammar-guided
decoding consistently achieves high validity while maintaining substantially lower exact sequence matching result than the device--pin representation, indicating reduced memorization at the
per-type level.

In contrast, the device--pin representation exhibits pronounced trade-offs between validity and
novelty when circuit-type tokens and renaming augmentation are applied, with performance varying
significantly across circuit types.
These per-type results further support the conclusion that the proposed bipartite formulation
enables robust, controllable, and non-memorizing topology generation across a wide range of
analog circuit classes.

\section{Dataset Statistics}
\label{app:dataset}

\begin{table}[h!]
\centering
\caption{Dataset Statistics}
\resizebox{\textwidth}{!}{%
\begin{tabular}{lccccccccccccccc|c}
\toprule
 & OpAmp & LDO & BGR & PConv & Oscil & General & Mirror & Mix & PAmp & PLL & Filt & Comp & VoltR & S Cap & A/D\_D/A & Total \\
\midrule
Raw netlist        
& 758 & 450 & 285 & 270 & 122 & 125 & 52 & 29 & 21 & 20 & 11 & 9 & 7 & 4 & 2 & 2165 \\
(\%)              
& 35.0 & 20.8 & 13.2 & 12.5 & 5.6 & 5.8 & 2.4 & 1.3 & 1.0 & 0.9 & 0.5 & 0.4 & 0.3 & 0.2 & 0.1 & 100 \\
\midrule
Augmented sequence 
& 132621 & 90000 & 55640 & 54000 & 22096 & 17322 & 7759 & 5154 & 3218 & 3787 & 1913 & 1800 & 1005 & 800 & 400 & 397515 \\
(\%)              
& 33.4 & 22.6 & 14.0 & 13.6 & 5.6 & 4.4 & 2.0 & 1.3 & 0.8 & 1.0 & 0.5 & 0.4 & 0.3 & 0.2 & 0.1 & 100 \\
\bottomrule
\end{tabular}%
}
\label{tab:dataset_statistics}
\end{table}

Table~\ref{tab:dataset_statistics} summarizes the composition of the dataset used in our
experiments, as introduced in Section~\ref{section:experimental}.
The dataset consists of 2,165 raw SPICE netlists spanning 14 circuit types and an additional
general category.

On this raw dataset, we apply sequence-level data augmentation strategies
described in Appendix~\ref{app:augmentation}, including the device renaming augmentation detailed
in Section~\ref{section:augmentation}.
These augmentations substantially increase the effective training set while preserving electrical
equivalence, resulting in a total of 397,515 augmented sequences used for model training.



\end{document}